\documentclass[a4paper,12pt,psamsfonts]{amsart}
\usepackage{amsmath}
\usepackage{amssymb}
\def\ifnextchar#1#2#3{\let\tmpnce=#1%
    \def\tmpnca{#2}\def\tmpncb{#3}\futurelet\tmpncc\ifnch}%
  \def\ifnch{\ifx\tmpncc\tmpnce\let\tmpncd=\tmpnca%
	\else\let\tmpncd=\tmpncb\fi\tmpncd}
\def\conjugate#1{\overline{#1}}

\def\d{\operatorname{d}\!}

\def\monthname{\ifcase\month\or Jan\or Feb\or March\or Apr\or %
    May\or June\or July\or Aug\or Sept\or Oct\or Nov\or Dec\fi}
\def\norm#1{\left\|#1\right\|}
      \def\normp#1_#2{\norm{#1}_{#2}}

\let\R\Real

\input epsf.tex  
\def\caption#1{\hfill \\
  \hbox{}\hfil{\footnotesize #1}\hfil\hbox{}}
\def\ifnextchar#1#2#3{\let\tmpnce=#1%
    \def\tmpnca{#2}\def\tmpncb{#3}\futurelet\tmpncc\ifnch}%
\def\ifnch{\ifx\tmpncc\tmpnce\let\tmpncd=\tmpnca%
    \else\let\tmpncd=\tmpncb\fi\tmpncd}
\def\tpeinture #1 by #2 (#3){
  \vtop to #2{
    \hrule width #1 height 0pt depth 0pt
    \vfill
    \epsfysize=#2 \epsfbox{#3}
    }
  }
\def\xpeinture #1 by #2 (#3){
  \hbox{$\vcenter to #2{
    \hrule width #1 height 0pt depth 0pt
    \vfill
    \epsfxsize=#1 \epsfbox{#3}
    }$}
  }
\def\ypeinture #1 by #2 (#3){
  \hbox{$\vcenter to #2{
    \hrule width #1 height 0pt depth 0pt
    \vfill
    \epsfysize=#2 \epsfbox{#3}
    }$}
  }
\def\peinture{\ypeinture}
\def\bpeinture #1 by #2 (#3){
  \vbox to #2{
    \hrule width #1 height 0pt depth 0pt
    \vfill
    \epsfysize=#2 \epsfbox{#3}
    }
  }
\def\Scaledpiu[#1] #2 by #3 (#4){{ %
   \dimen0=#2 \dimen1=#3 %
   \if#1t
       \tpeinture \dimen0 by \dimen1 (#4)%
   \else\if#1b
        \bpeinture \dimen0 by \dimen1 (#4)%
        \else\if#1x
             \xpeinture \dimen0 by \dimen1 (#4)%
             \else
             \ypeinture \dimen0 by \dimen1 (#4)%
            \fi\fi
    \fi}}
\def\Scaledpiv #1 by #2 (#3){{%
   \dimen0=#1 \dimen1=#2%
   \peinture \dimen0 by \dimen1 (#3) %
   }}
\def\scaledpicture{\ifnextchar[{\Scaledpiu}{\Scaledpiv}}
\def\centredpicture #1 by #2 (#3){
   \par\centerline{\hbox{
   \scaledpicture #1 by #2 (#3)}
   }}

     \input epsf.tex
\title[Biconcave Axisymmetric Vesicles]{Conditions for the Formation of Axisymmetric Biconcave Vesicles}

\author[Thomas Au]{Thomas Kwok-keung Au}
\address{Department of Mathematics, The Chinese University of Hong Kong}
\email{thomasau@cuhk.edu.hk}

\author[Tom Wan]{Tom Yau-heng Wan}
\address{Department of Mathematics, The Chinese University of Hong Kong}
\email{tomwan@cuhk.edu.hk}

\def\HelF{{\mathcal F}}
\def\tlambda{\tilde{\lambda}}
\def\tp{\tilde{p}}
\def\bX{{\mathbf X}}

\begin{document}
\maketitle

\setlength{\baselineskip}{24pt}

\section*{Introduction}
\setcounter{section}{0}
The extraordinary biconcave shape of a red blood cell has attracted much interest for many years.  In the last two decades, people generally accepted the shape of biological membranes such as blood cells is closely related to the formation of lipid bilayer vesicle in aqueous medium.  Based on the elasticity of lipid bilayers proposed by Helfrich \cite{Helfrich1973}, the shape $\Sigma$, regarded as an embedded surface in ${\mathbb R}^3$, is determined by the minimum of the bending energy involving the volume $\text{V}(\Sigma)$ enclosed by $\Sigma$, the area $\text{A}(\Sigma)$, the mean curvature $H$ and the Gaussian curvature $K$ of $\Sigma$.  More precisely, Helfrich suggested to study the bending energy 
$$
\frac{1}{2}k_c\oint_\Sigma (2H+c_0)^2 \d A + \frac{1}{2}\conjugate{k_c}\oint_\Sigma K \d A + \lambda \text{A}(\Sigma) + p\text{V}(\Sigma),
$$
where $k_c$, $\bar{k}_c$, $c_0$, $\lambda$, and $p$ are constants interpreted as follow: $k_c$ is the bending rigidity, $\bar{k}_c$ the Gaussian curvature modulus, $c_0$ the spontaneous curvature, $\lambda$ the tensile stress, and $p=p_o-p_i$ the osmotic pressure difference between the outer ($p_o$) and inner ($p_i$) media.  Here, we have taken a geometric sign convention so that Helfrich's original bending energy should be written in the above form.

According to the Gauss-Bonnet Theorem, the second integral in the bending energy is a topological constant.  Therefore, within a certain topological class of $\Sigma$, it is sufficient to study the functional,
$$
\HelF(\Sigma) = \oint_\Sigma (2H+c_0)^2 \d A + \tlambda \text{A}(\Sigma) + \tp\text{V}(\Sigma)
$$
where $c_0$, $\tlambda=2\lambda/k_c$, and $\tp = 2p/k_c$ are the constant parameters of the functional.  This functional will be referred as the Helfrich functional in this article.

Many interesting surfaces such as minimal surfaces, constant mean curvature surfaces and Willmore surfaces can be regarded as critical points of the Helfrich functional for suitable combinations of the parameters $c_0$, $\tlambda$, and $\tp$. New axisymmetric explicit solutions had also been found recently \cite{Naito-Okuda-OuYang2}.

When $c_0=\tlambda=\tp=0$, the functional $\HelF(\Sigma)$ is referred as Willmore functional in differential geometry which has been widely studied in recent decades.  Moreover, there are some important analysis and open problems concerning the Willmore functional, \cite[ch.~7]{Willmore}.  As it is observed by physicists \cite{Helfrich1976}, Willmore functional is not a good model for the shape of red blood cells.  This fact can also be seen by the result of geometers that the unique minimum of the Willmore functional for topologically spherical vesicles (embedded surfaces of genus zero in terms of topologist) is the round sphere. Therefore, not all combinations of the parameters give stationary vesicles of shape similar to red blood cells observed experimentally.

On the other hand, a typical way to investigate the minimizing surface is by finding solutions to the variational equation of the functional $\HelF$.   A large class of axisymmetric stationary vesicles of spherical and toroidal topology has been calculated, \cite{Helfrich1976,Luke,Seifert,Mutz-Bensimon,OuYang}.  In these works, among other interesting shapes of lipid bilayer vesicles, the shape of the red blood cell can be simulated numerically or given by a special solution a specific variational equation with suitable combinations of the parameters.  It is natural to ask for the conditions on the parameters such that the Helfrich functional possesses a stationary vesicle of biconcave shape.

\subsection*{Main Results}
In this article, we are going to give clear conditions on how biconcave axisymmetric surfaces are formed.  More precisely, we find a sufficent condition for that the Helfrich shape equation of axisymmetric vesicles to have solutions of biconcave shape.  Besides, we exhibit that when the equation has a solution with reflection symmetry or of biconcave shape, certain geometric quantities of the vesicle must obey some conditions governed by the parameters.
These conditions on the parameters are very mild.  In particular, the case that $c_0>0$, $\tlambda >0$, and $\tp >0$ is sufficient to ensure the existence of biconcave  solution.  We also briefly comment on how the combination of the parameters $c_0$, $\tlambda$, and $\tp$ affects the existence of biconcave solution.

The sufficient condition for the formation of biconcave vesicle may be formulated in terms of a cubic polynomial,
\begin{equation}\label{eqn-Q}
Q(t) = t^3 + 2c_0 t^2 + (c_0^2 + \tlambda)t - \frac{\tp}{2}.
\end{equation}
We proved that {\bf if all roots of $Q(t)$ are positive, then there is always an axisymmetric biconcave vesicle which is the stationary surface for the Helfrich functional}.  This result will be summary by the table at the end of this section (p.~\pageref{summarytable}) along with some typical pictures of $Q(t)$ and the corresponding graphs of the solution and its derivative.

It is easy to verify that $Q(t)\le - \frac{\tp}{2}$ for all $t\le 0$ if $c_0>0$, $\tlambda>0$, and $\tp>0$ and hence all roots are positive. Mathematically, that all roots of $Q(t)$ are positive can be written as
$$
\max\left\{Q(t):-\infty < t \leq 0\right\} < 0.
$$
Explicit formula for this in terms of the parameters can also be found. However, the above is more concise and precise enough. 

For the necessary condition, we observe that if $c_0>0$ and there is an axisymmetric stationary vesicle of biconcave shape, then its curvature at the center must be smaller than the first positive root of $Q(t)$. In fact, we have
$$
2c_0{w_0'}^2+ (c_0^2 + \tlambda)w_0' - \frac{\tp}{2}<0,
$$
where $w_0'$ is the meridinal curvature (or ${w_0'}^2$ is the Gaussian curvature) at the center.  

Furthermore, if the axisymmetric vesicle has a reflection symmetry with respect to a plane perpendicular to the rotation axis, then we obtain a relation bewteen the Gaussian curvature at the ``equator'' (the intersection circle of the axisymmetric vesicle with the plane of reflection) and the radius $r_\infty$ of the ``equator" in terms of the polynomial $Q(t)$ given by (\ref{eqn-Q}) as follows. 
$$
K(r_\infty)^2 = \frac{-1}{r_\infty} Q\left(\frac{-1}{r_\infty}\right).
$$

This paper is organized as follows.  In \S 1, we rewrite the Helfrich shape equation of axisymmetric vesicles into two useful forms that are more convenient for our later discussion.  In \S 2, the change of the principal curvature in meridinal direction is analyzed and a necessary condition for the existence of biconcave surface is given.  In \S 3, we present a sufficient condition for the existence of biconcave surface.    In \S 4, we give a brief discussion of other situation such as axisymmetric vesicles which do not have reflection symmetry or which is not biconcave.

We would like to thank K.~S.~Chou for his discussions and valuable suggestions on the writing of this manuscript.

\clearpage

\label{summarytable}
\noindent
\begin{tabular}{c||c}
Graph of $Q$ and positions of $w_0'$ &Graphs of solution $z$ and graphs of $w=z'$ \\
\hline\hline
\parbox{7cm}{
\begin{center}
$c_0 = 1$, $\tlambda = 0.25$, $\tp = 1$
\end{center}
\xpeinture 7cm by 4.5cm (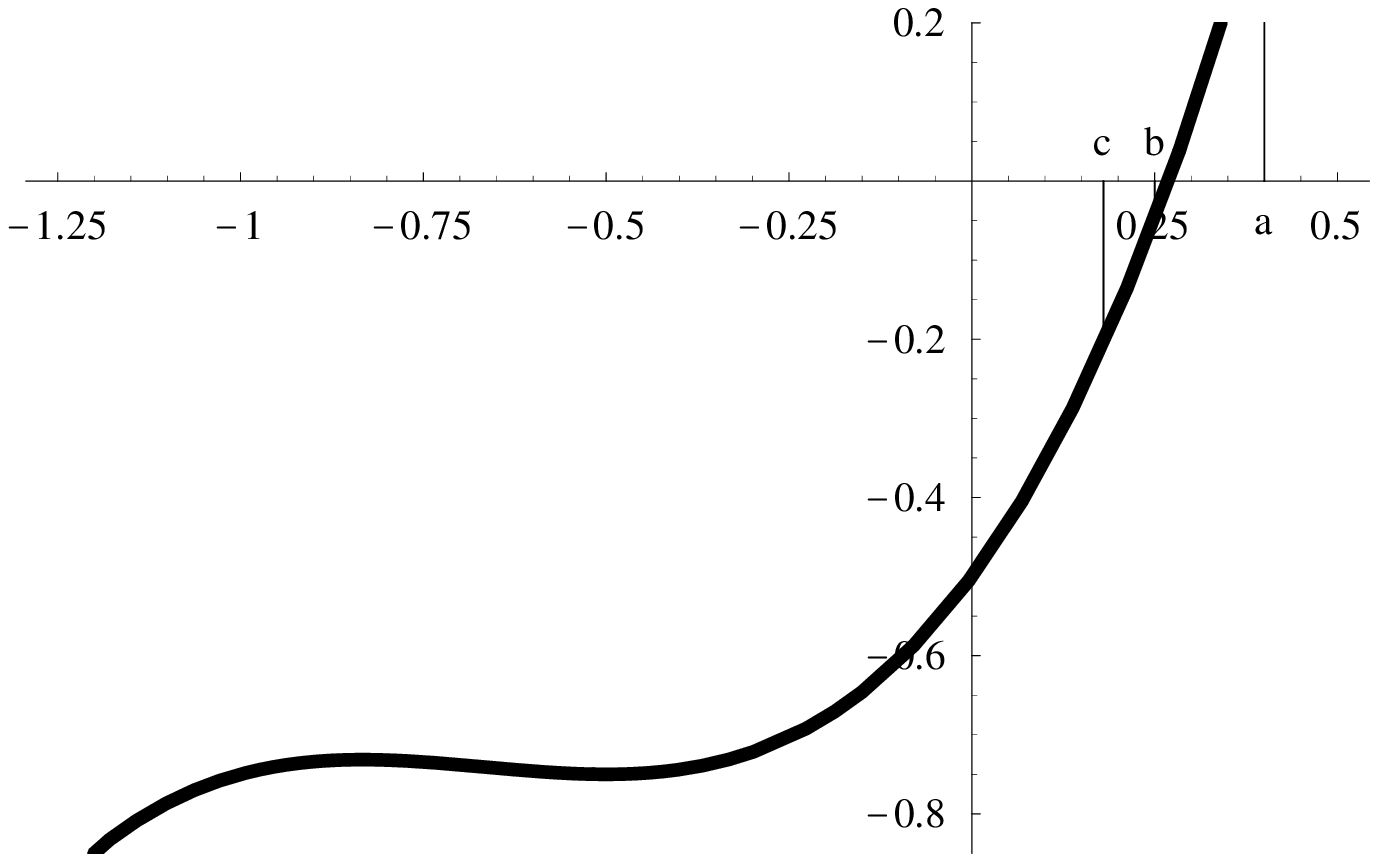)}
&\begin{tabular}{cc}
\multicolumn{2}{l}{\footnotesize (a) $w_0'=0.4$} \\
\parbox{4cm}{\xpeinture 4cm by 2.5cm (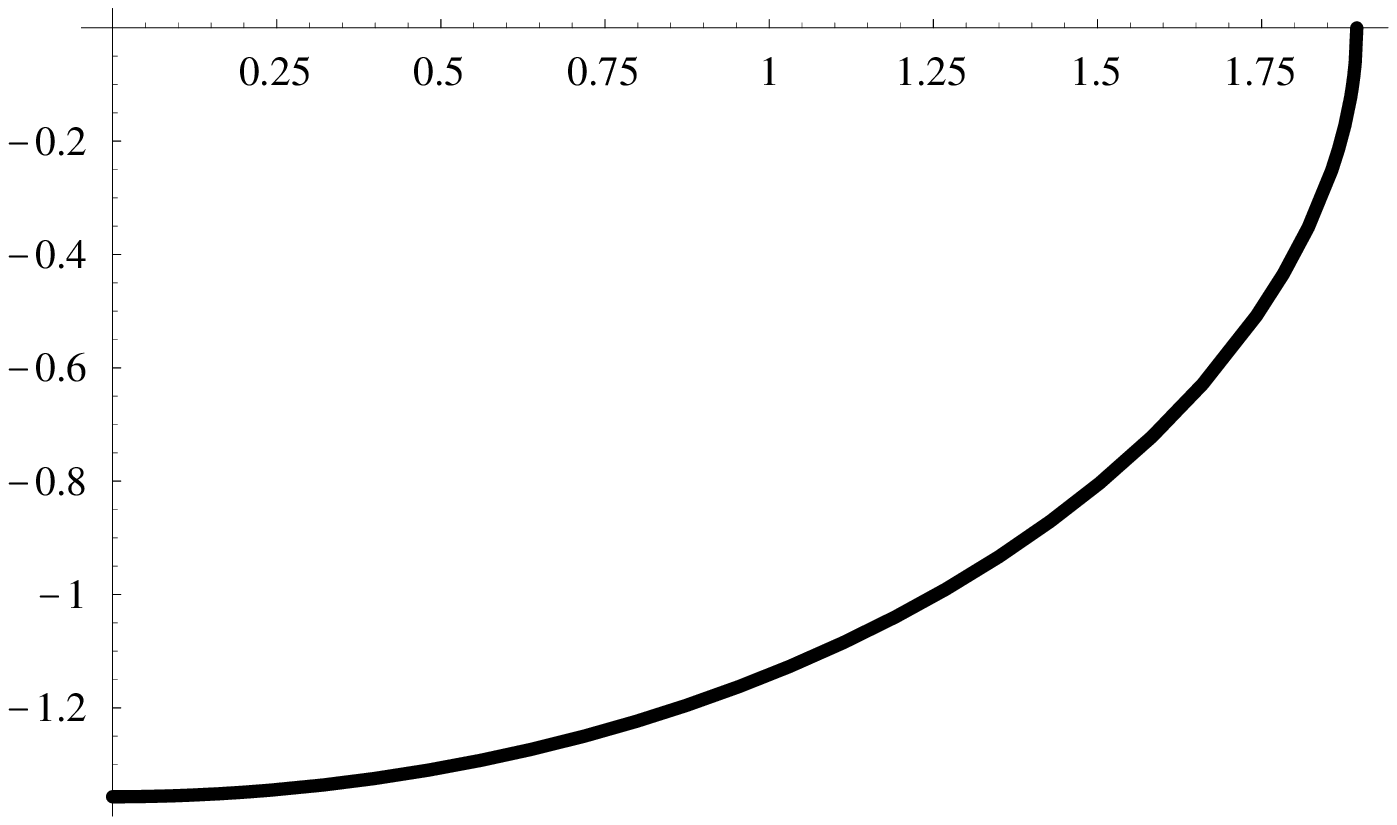)}
&\parbox{4cm}{\xpeinture 4cm by 2.5cm (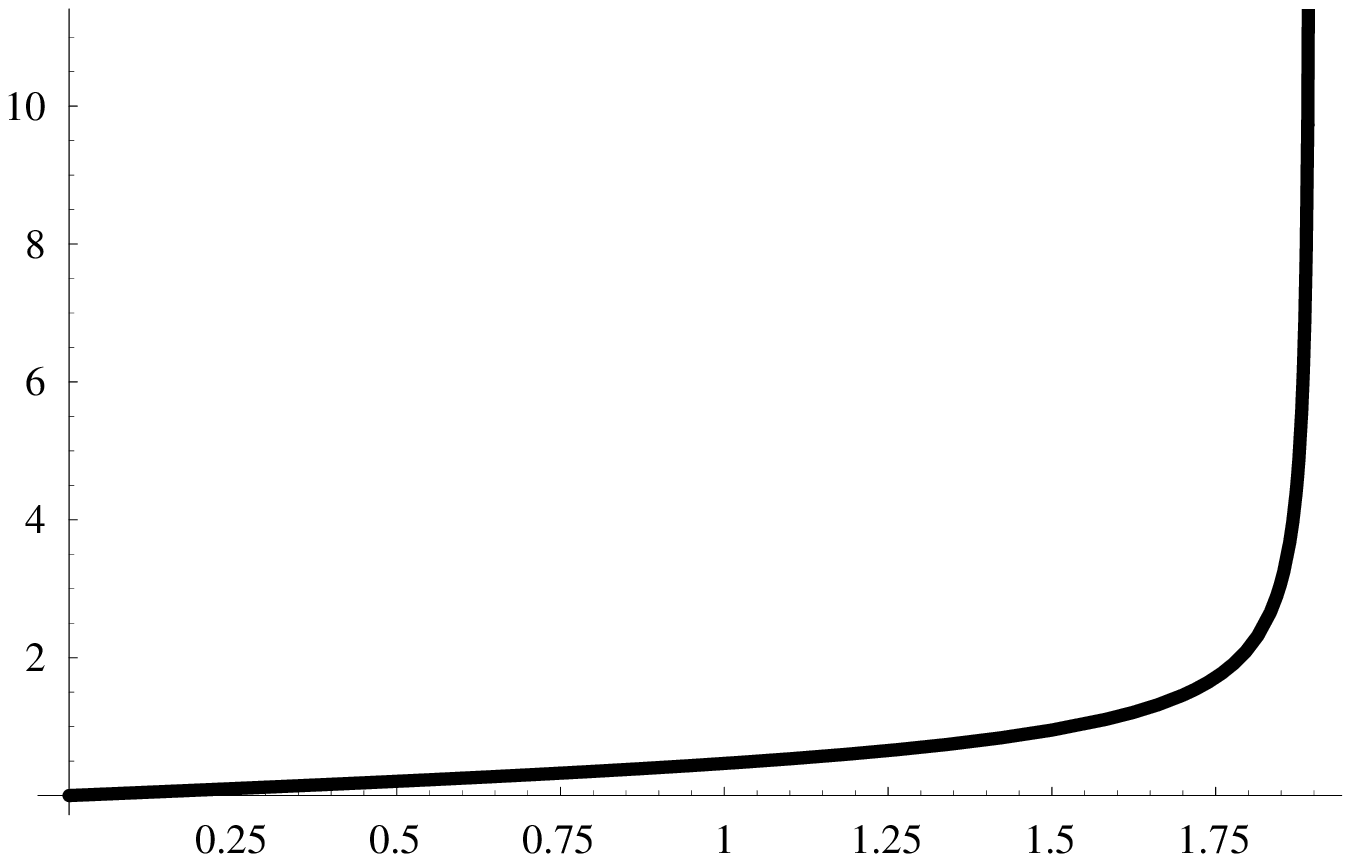)} \\ 
\multicolumn{2}{l}{\footnotesize (b) $w_0'=0.25$} \\
\parbox{4cm}{\xpeinture 4cm by 2.5cm (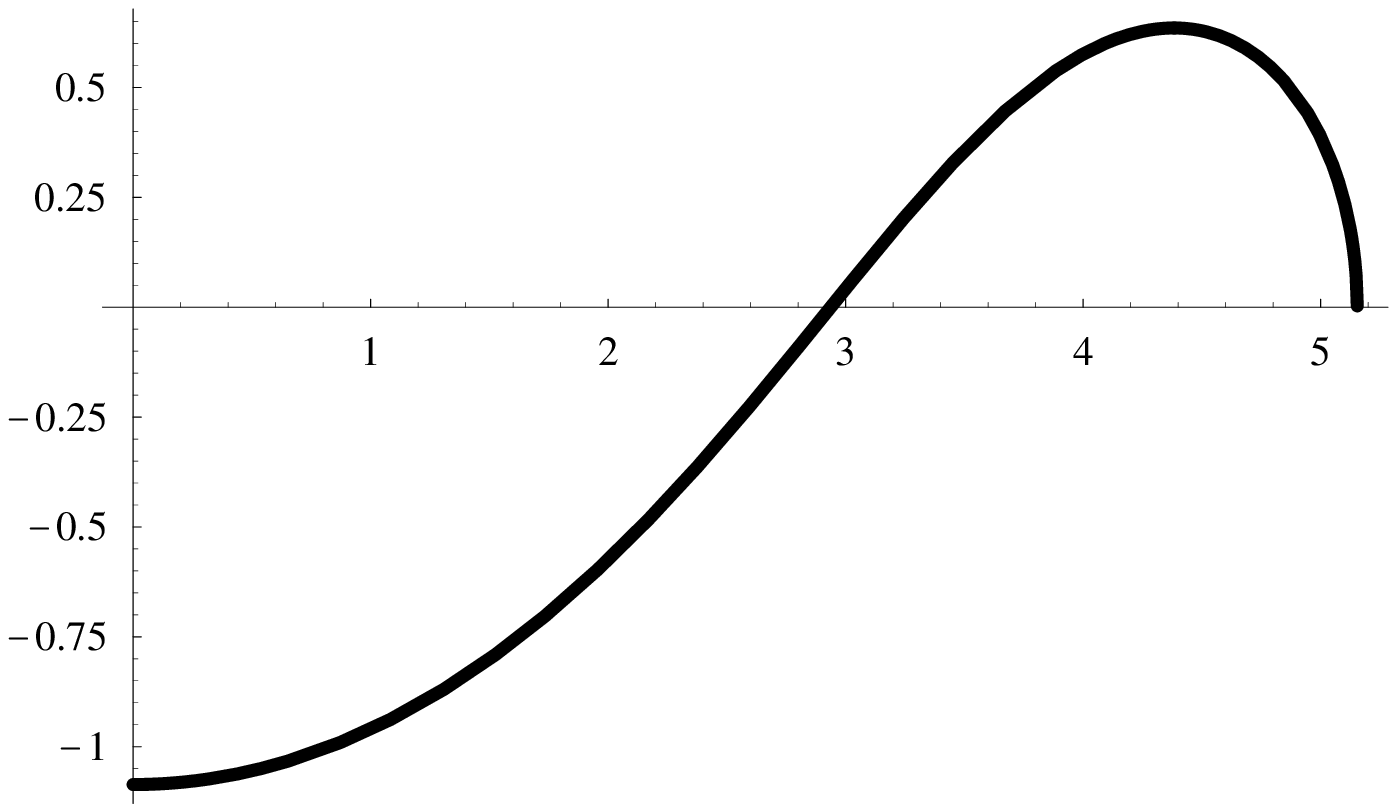)}
&\parbox{4cm}{\xpeinture 4cm by 2.5cm (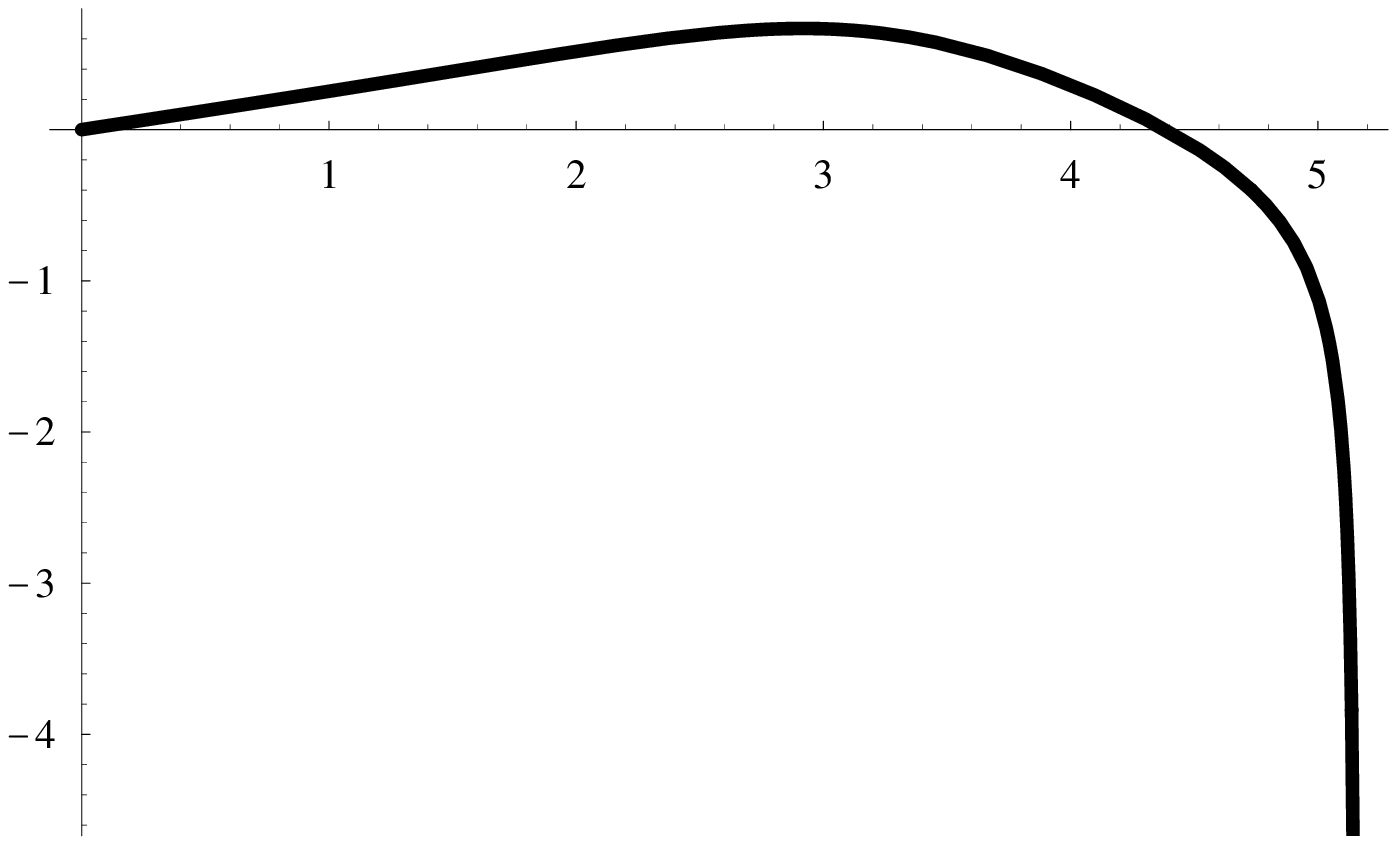)} \\
\multicolumn{2}{l}{\footnotesize (c) $w_0'=0.18$} \\
\parbox{4cm}{\xpeinture 4cm by 2.5cm (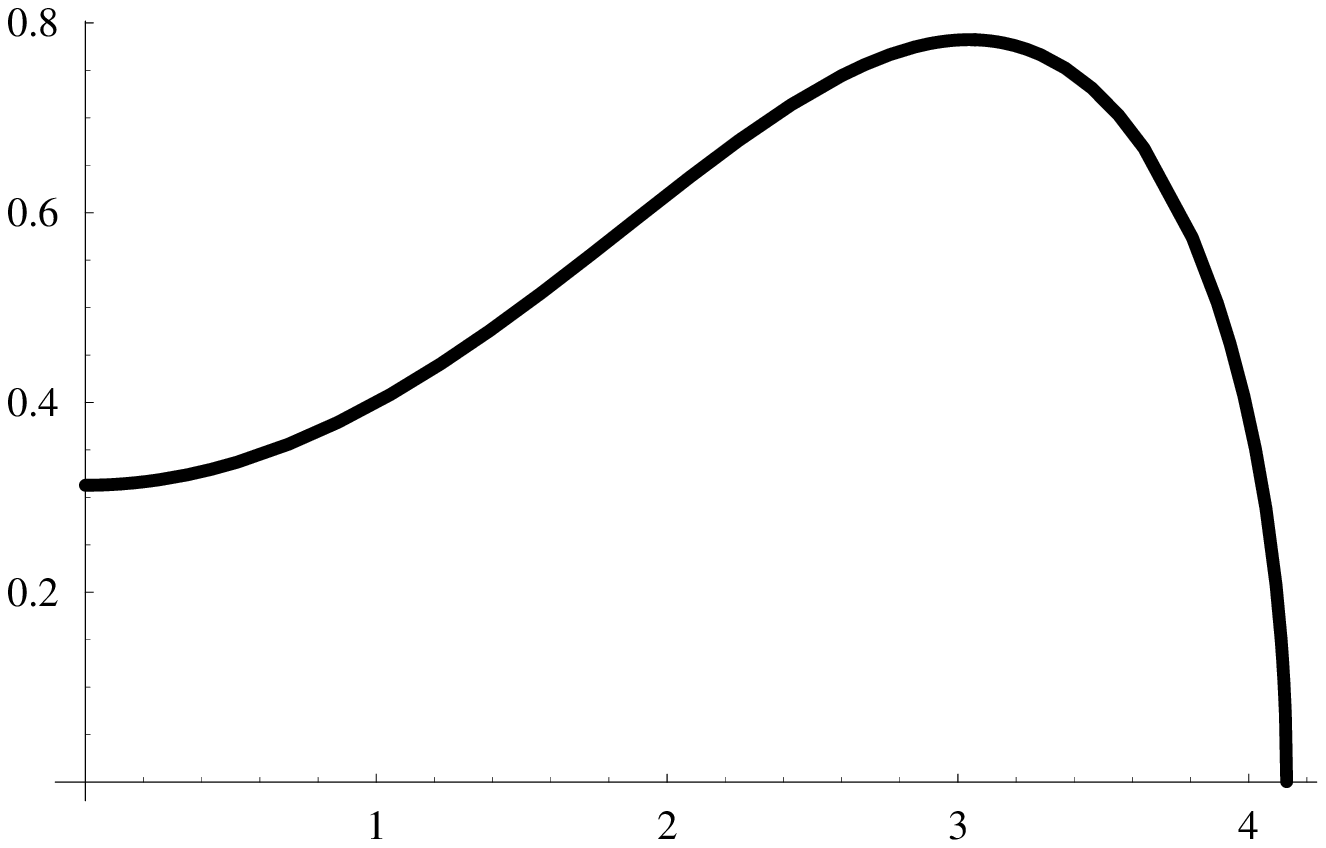)}
&\parbox{4cm}{\xpeinture 4cm by 2.5cm (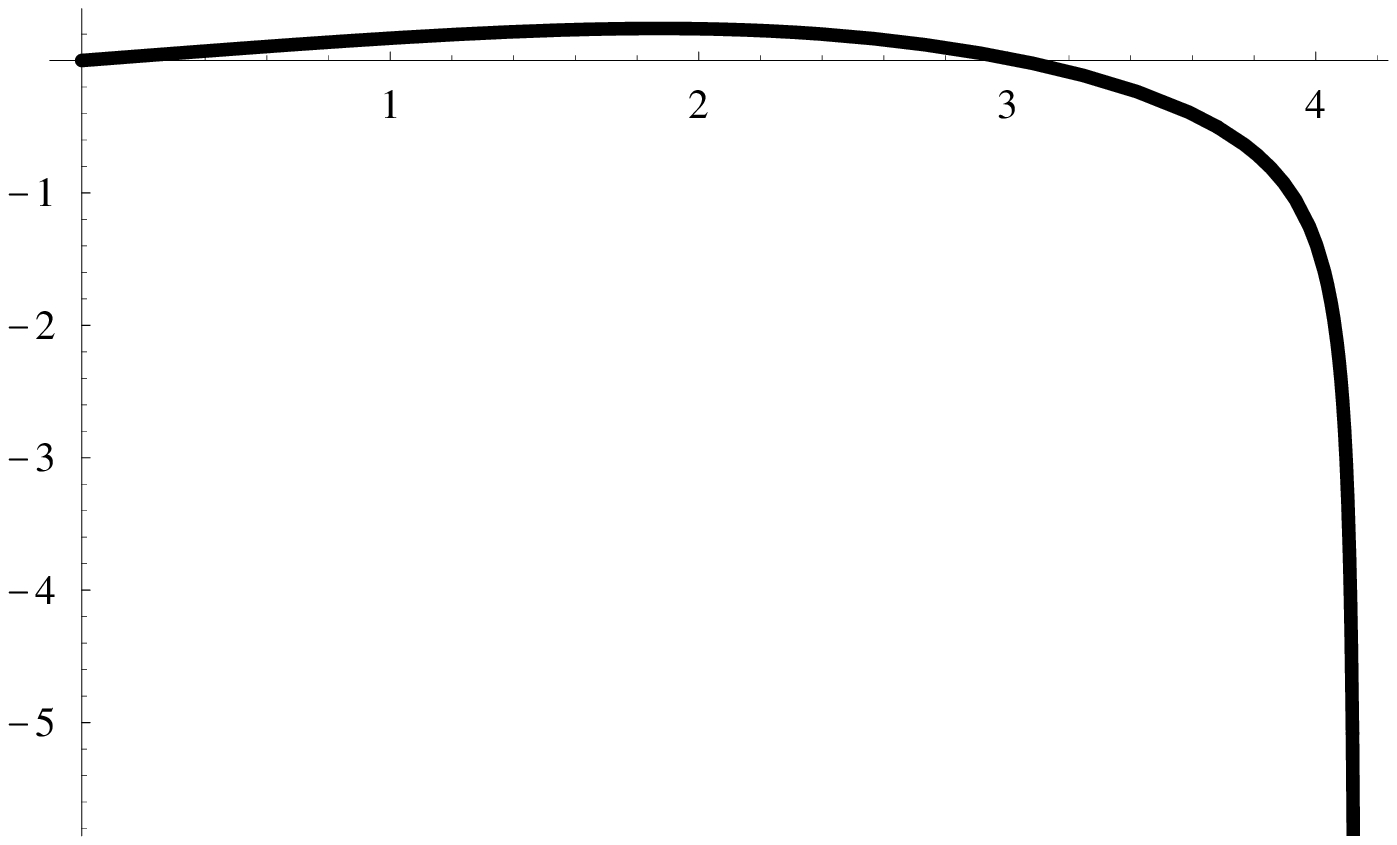)}
 \end{tabular}\\
\hline\hline
\end{tabular}

\medskip

\noindent
\begin{tabular}{c||c}
Graph of $Q$ and positions of $w_0'$ &Graphs of solution $z$ and graphs of $w=z'$ \\
\hline\hline
\parbox{7cm}{
\begin{center}
$c_0 = 1$, $\tlambda = -0.3$, $\tp = 1$
\end{center}
\xpeinture 7cm by 4.5cm (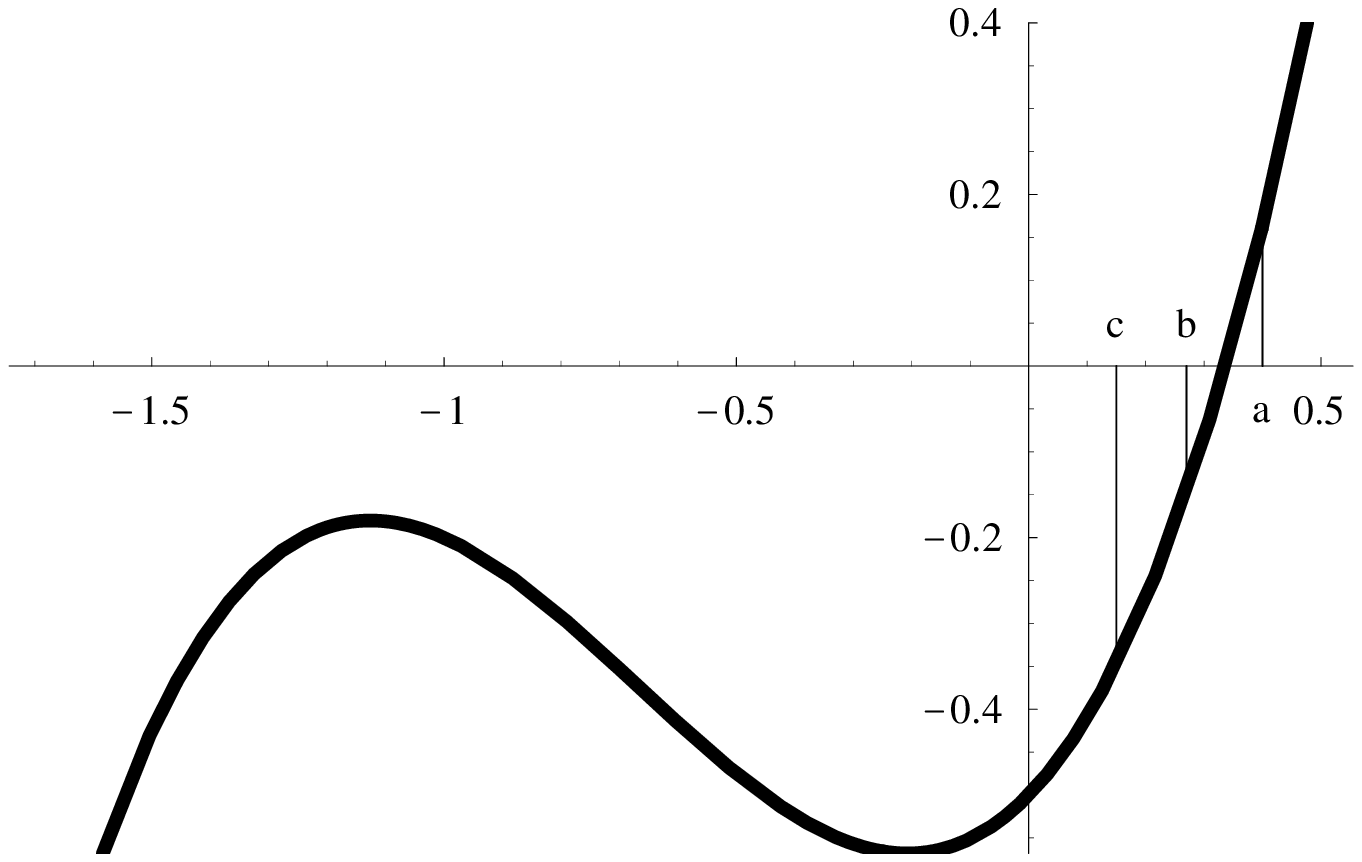)}
&\begin{tabular}{cc}
\multicolumn{2}{l}{\footnotesize (a) $w_0'=0.4$} \\
\parbox{4cm}{\xpeinture 4cm by 2.5cm (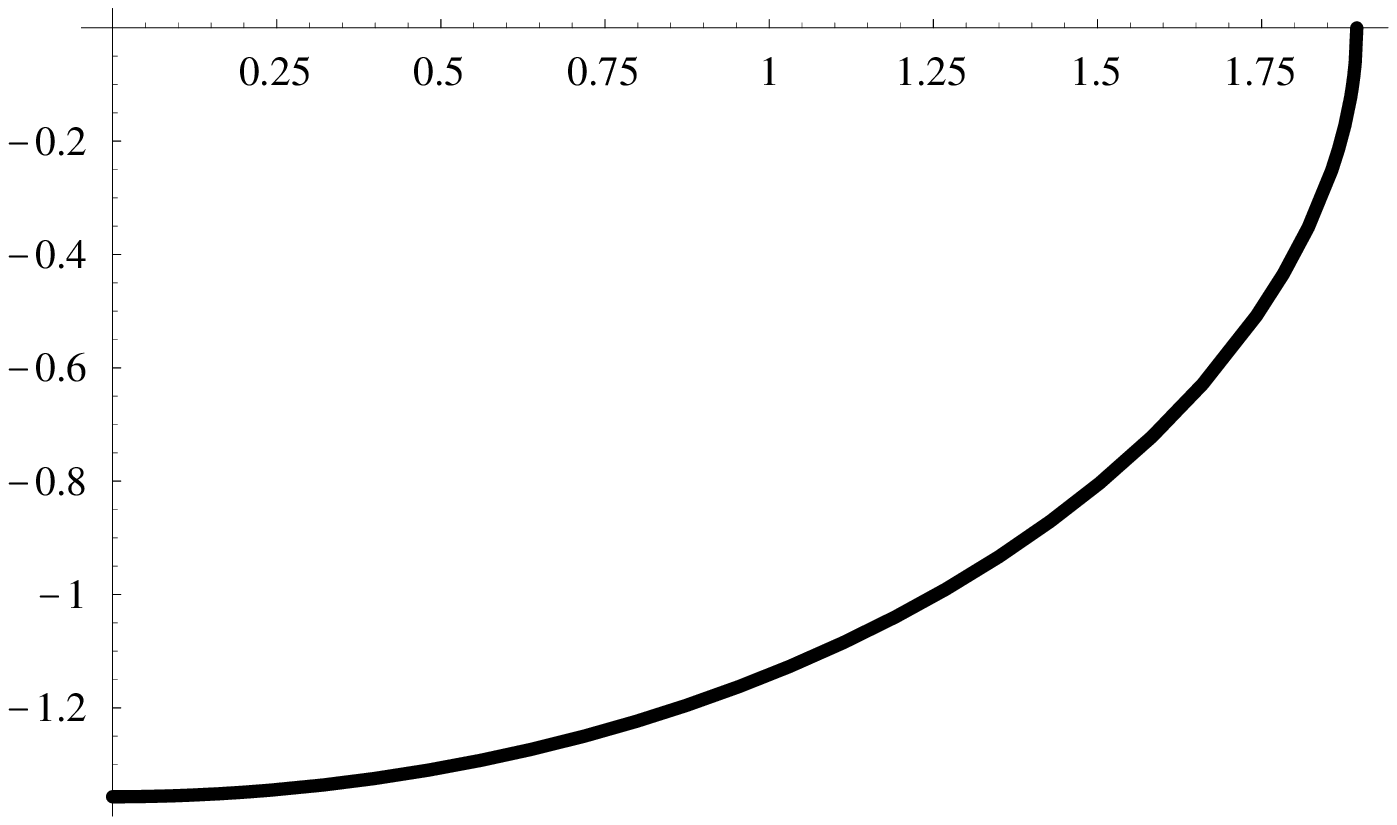)}
&\parbox{4cm}{\xpeinture 4cm by 2.5cm (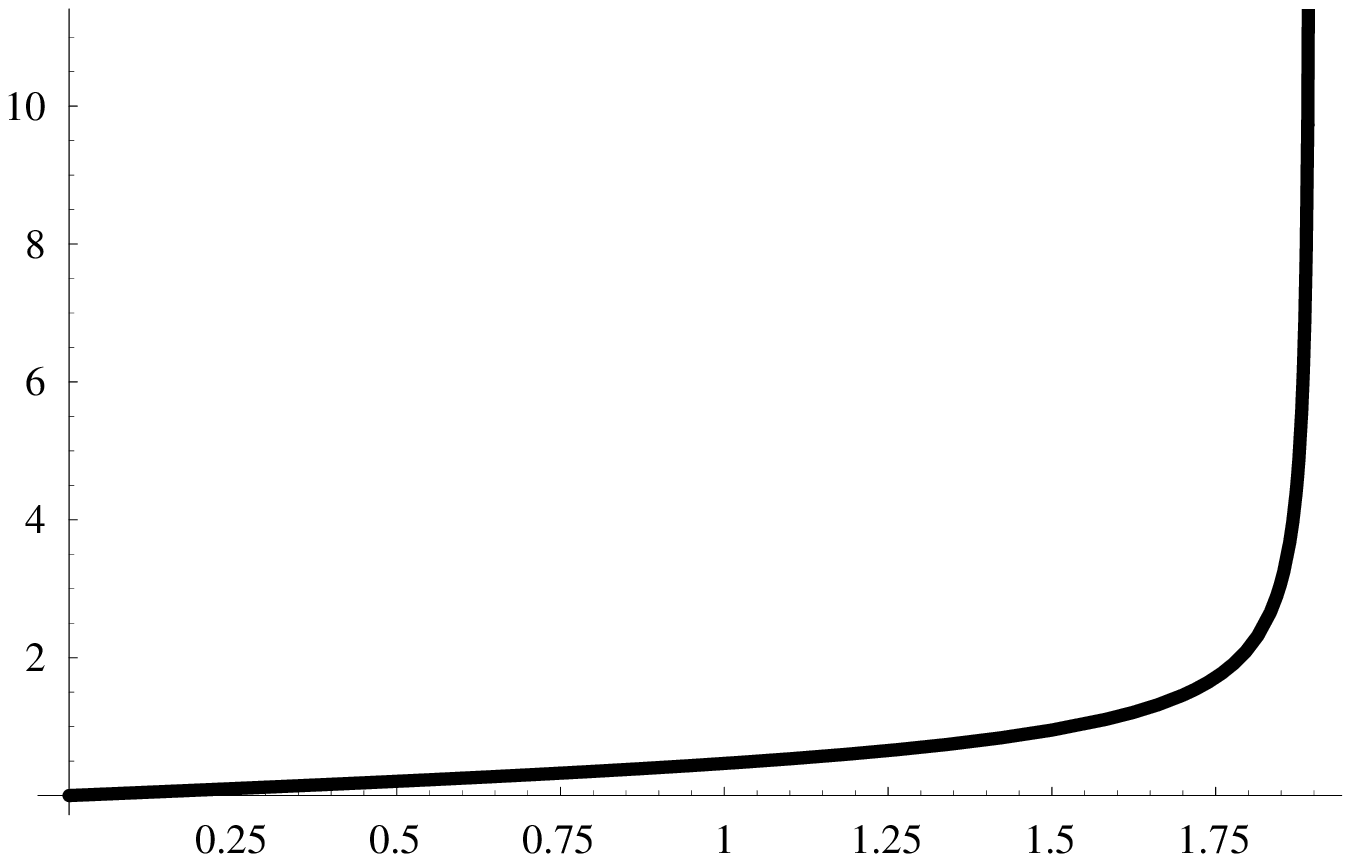)} \\
\multicolumn{2}{l}{\footnotesize (b) $w_0'=0.27$} \\
\parbox{4cm}{\xpeinture 4cm by 2.5cm (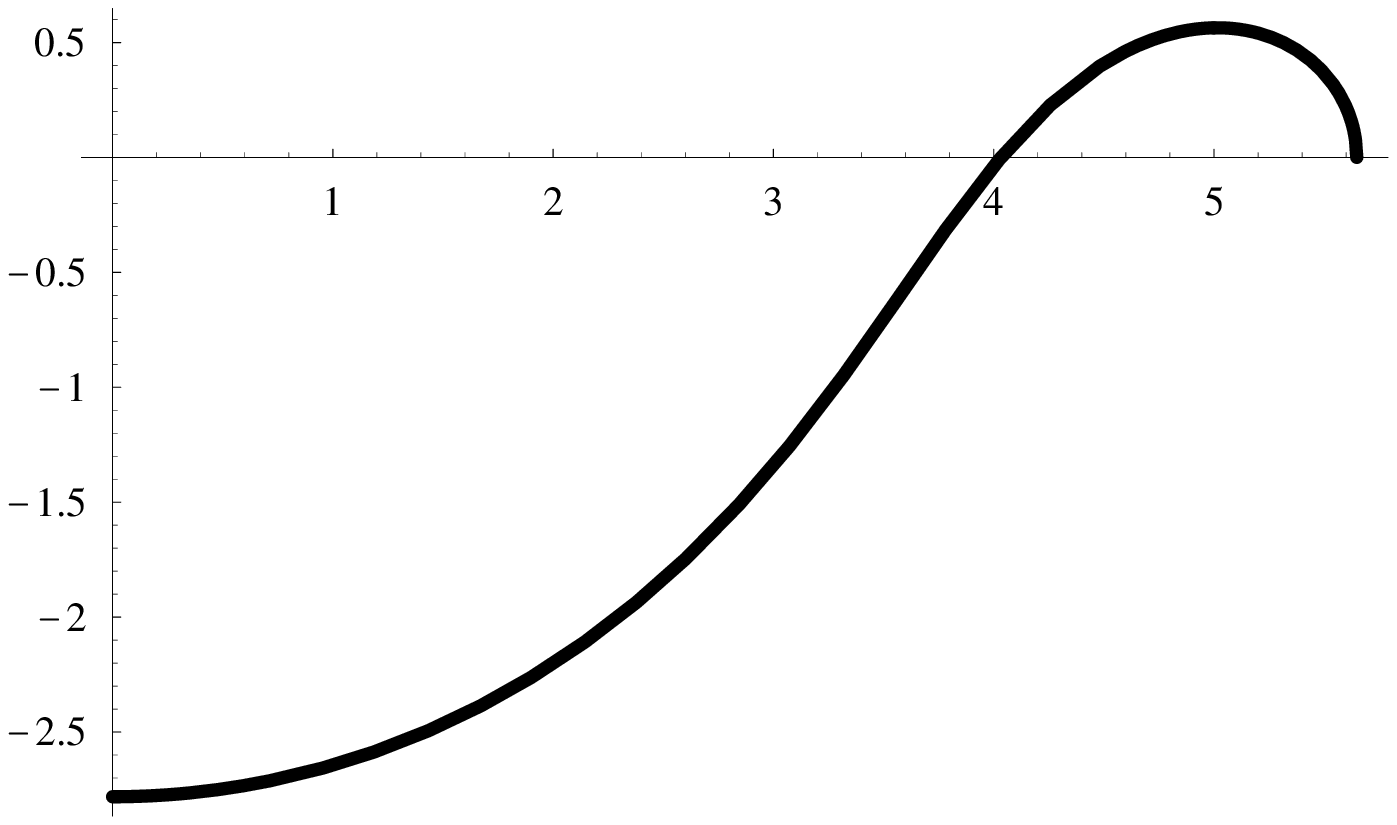)}
&\parbox{4cm}{\xpeinture 4cm by 2.5cm (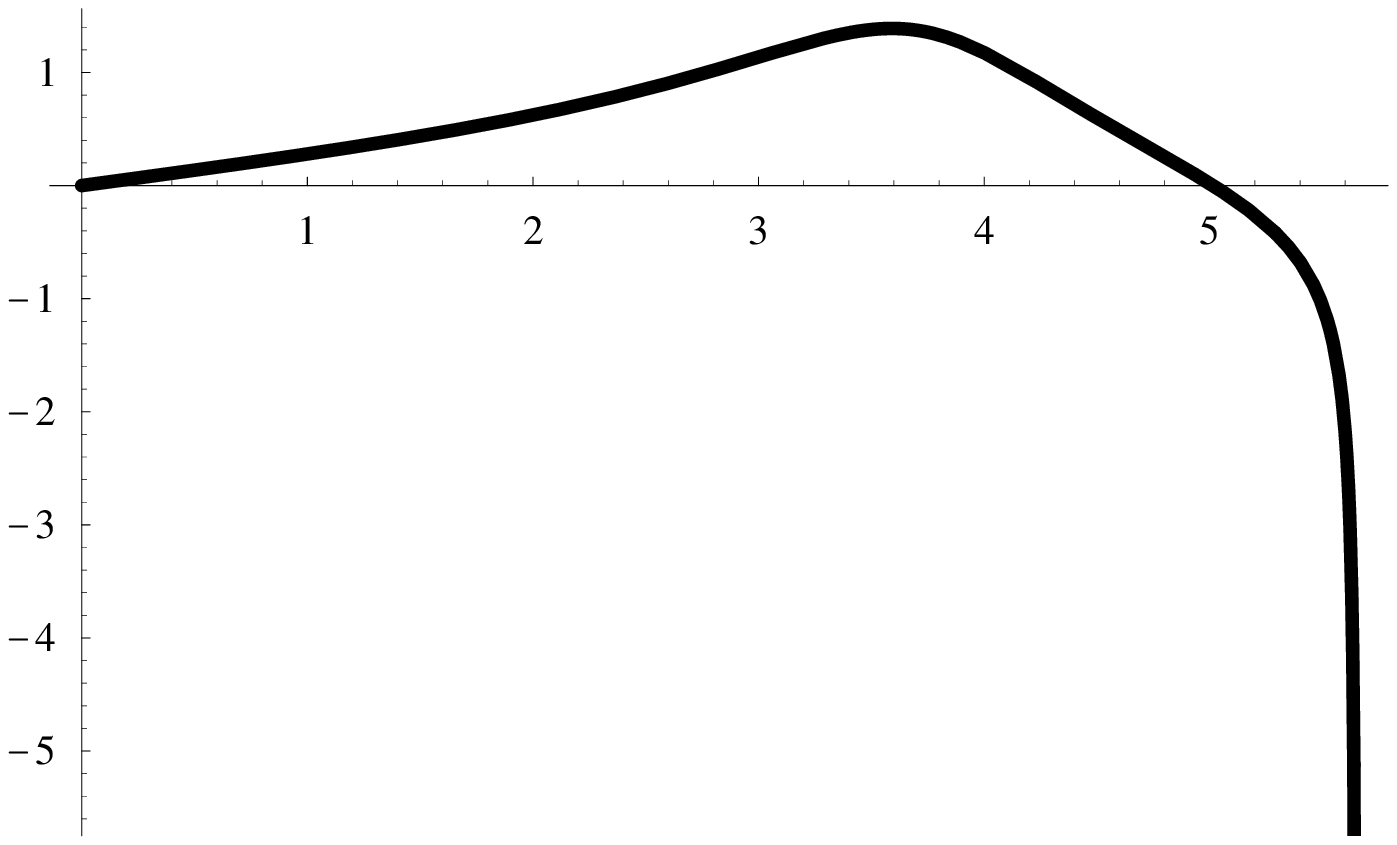)} \\
\multicolumn{2}{l}{\footnotesize (c) $w_0'=0.15$} \\
\parbox{4cm}{\xpeinture 4cm by 2.5cm (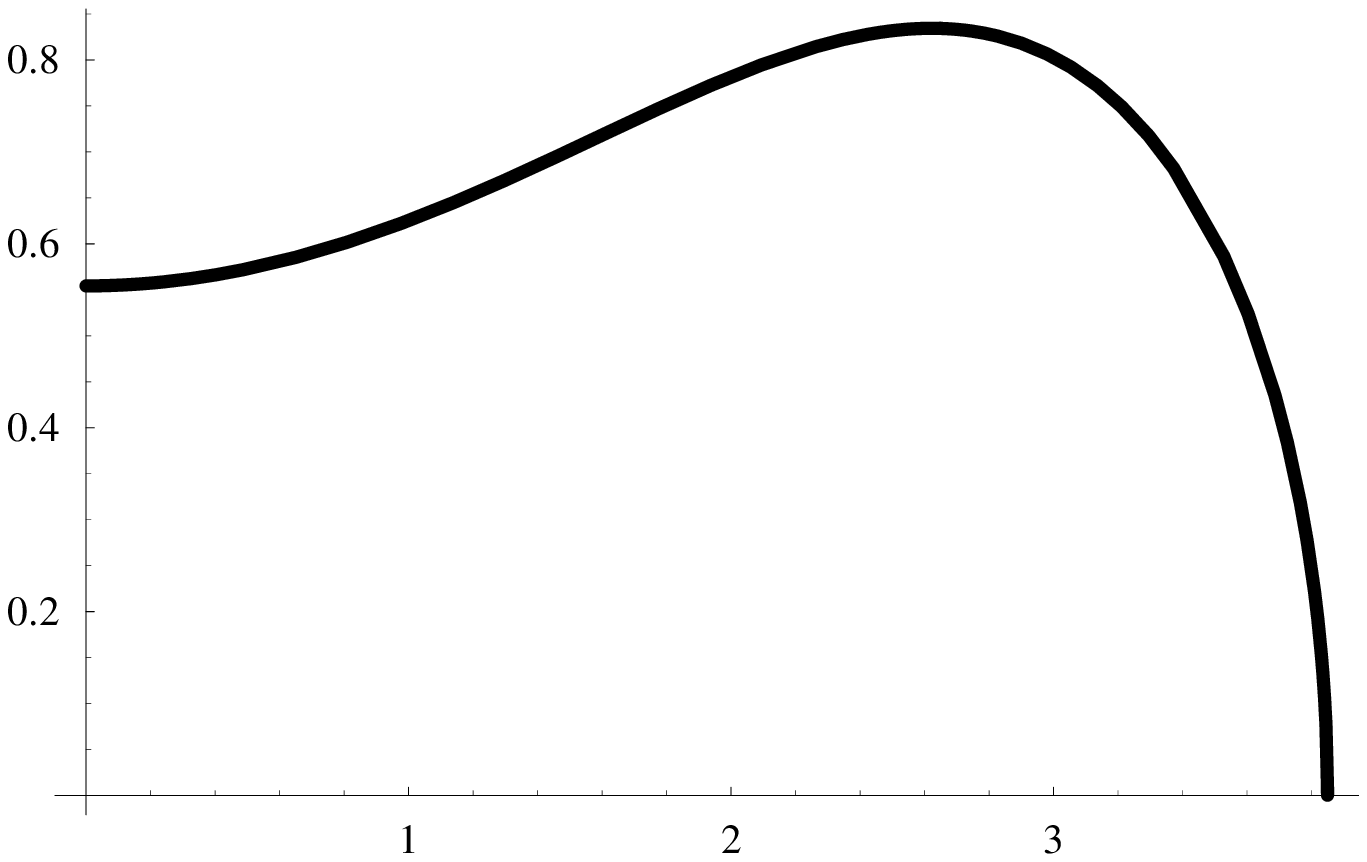)}
&\parbox{4cm}{\xpeinture 4cm by 2.5cm (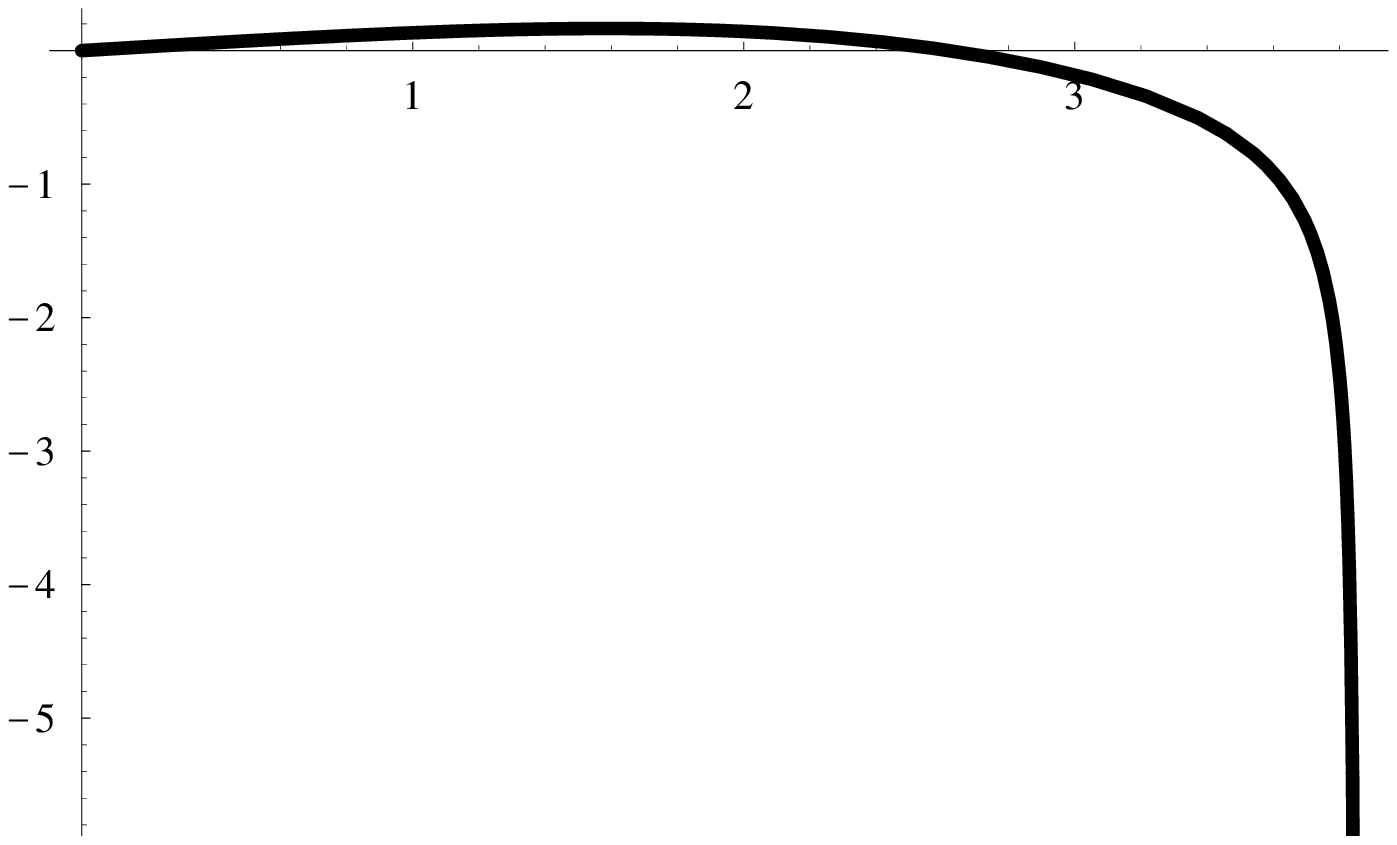)}
 \end{tabular}\\
\hline\hline
\end{tabular}

\medskip

\noindent
\begin{tabular}{c||c}
Graph of $Q$ and positions of $w_0'$ &Graphs of solution $z$ and graphs of $w=z'$ \\
\hline\hline
\parbox{7cm}{
\begin{center}
$c_0 = -1$, $\tlambda = -0.25$, $\tp = 1$
\end{center}
\xpeinture 7cm by 4.5cm (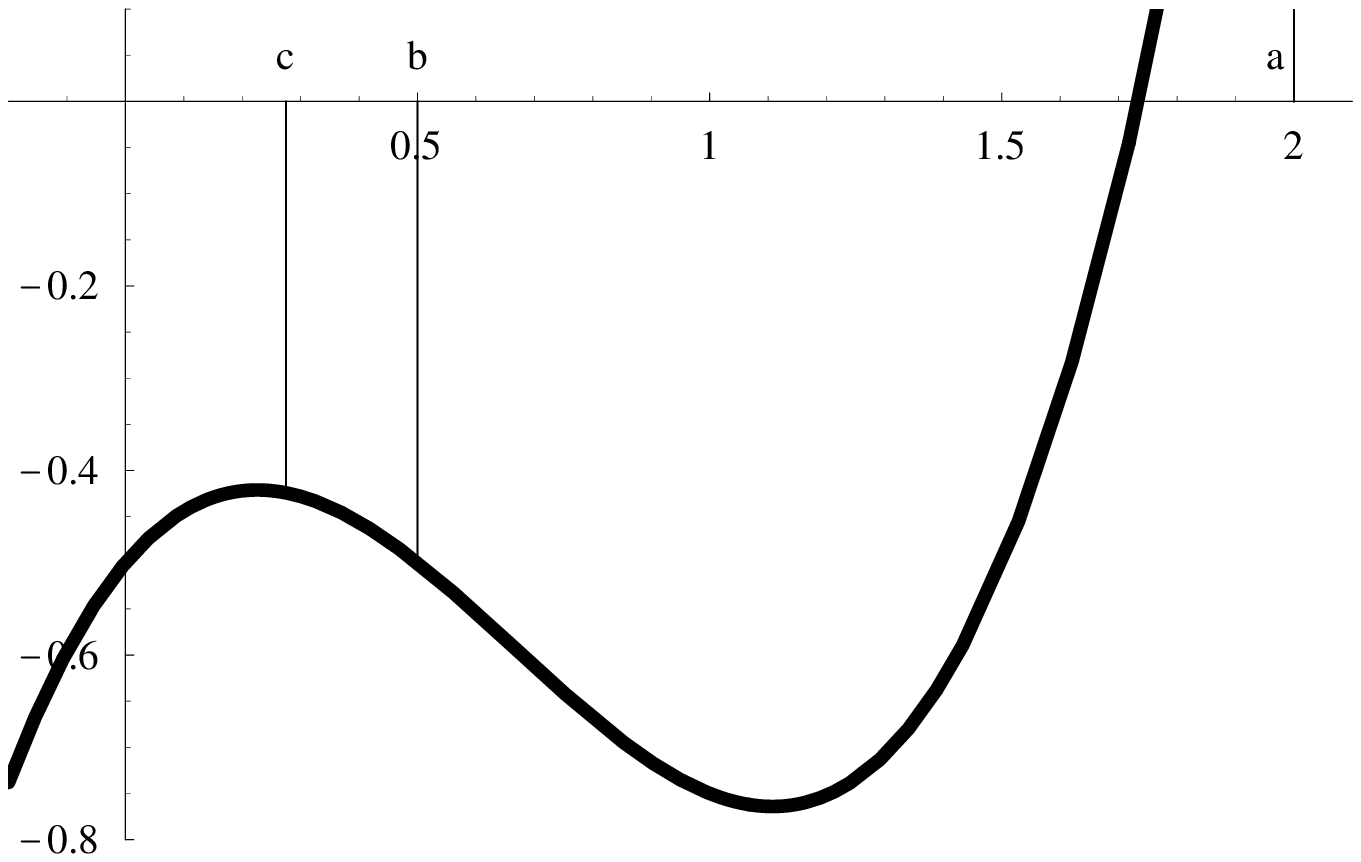)}
&\begin{tabular}{cc}
\multicolumn{2}{l}{\footnotesize (a) $w_0'=2$} \\
\parbox{4cm}{\xpeinture 4cm by 2.5cm (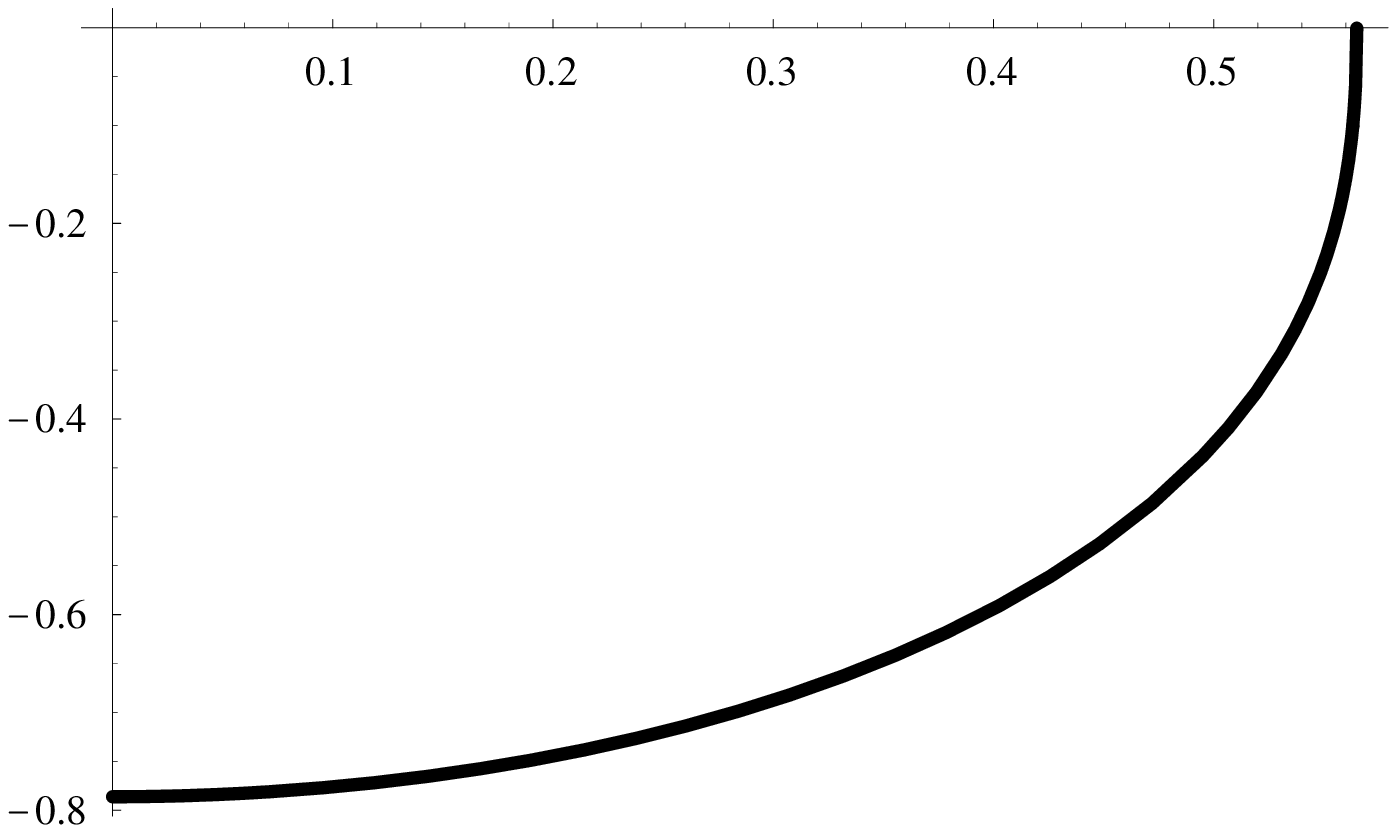)}
&\parbox{4cm}{\xpeinture 4cm by 2.5cm (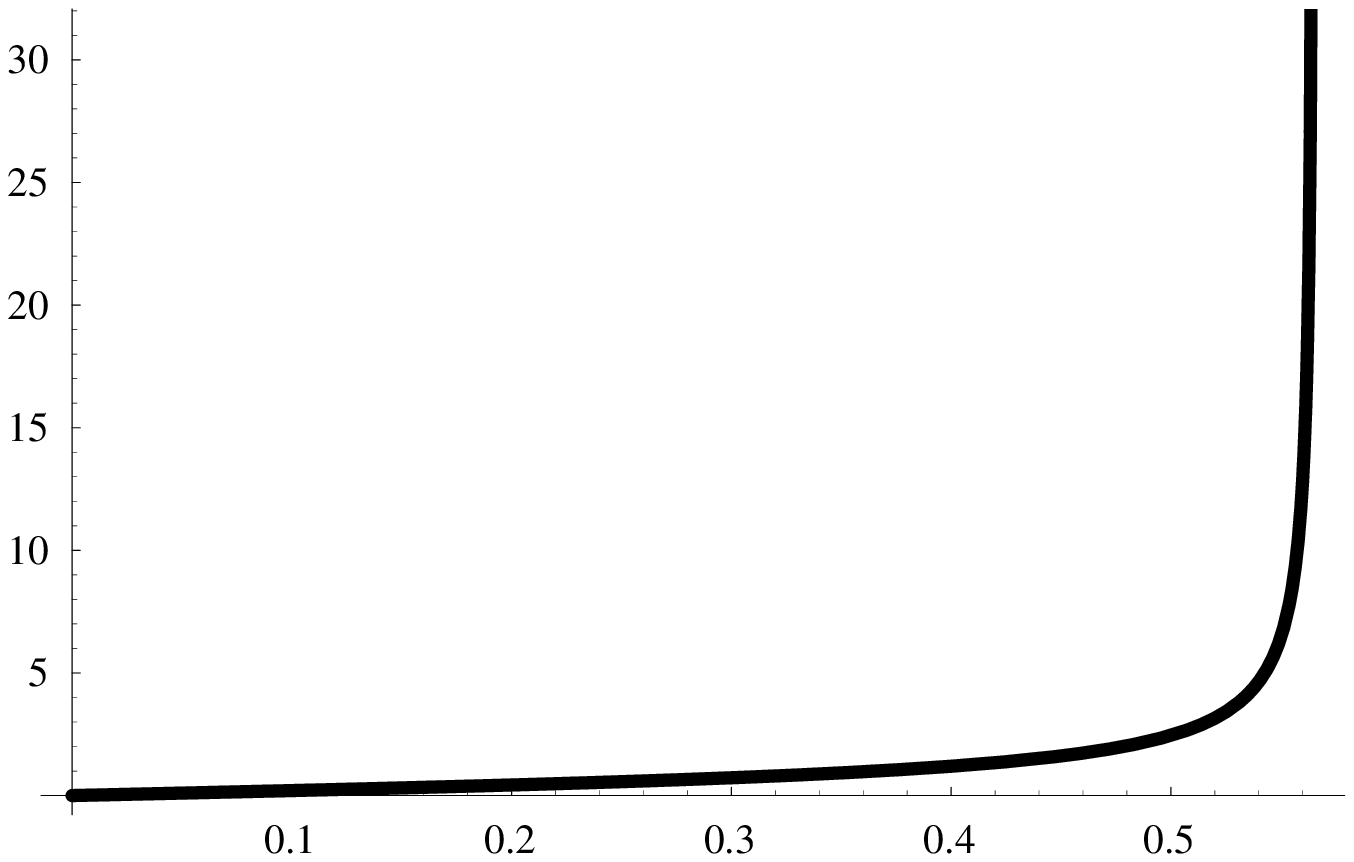)} \\
\multicolumn{2}{l}{\footnotesize (b) $w_0'=0.5$} \\
\parbox{4cm}{\xpeinture 4cm by 2.5cm (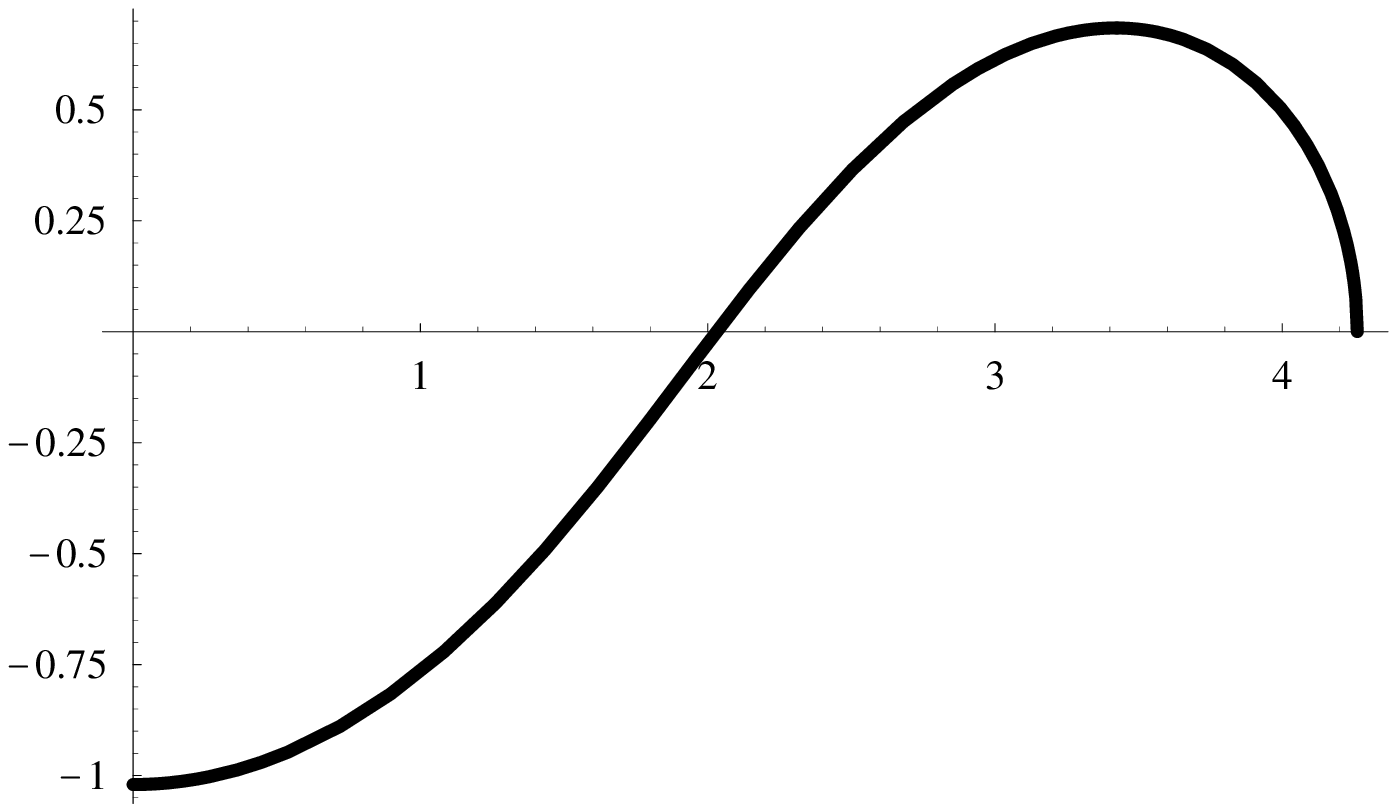)}
&\parbox{4cm}{\xpeinture 4cm by 2.5cm (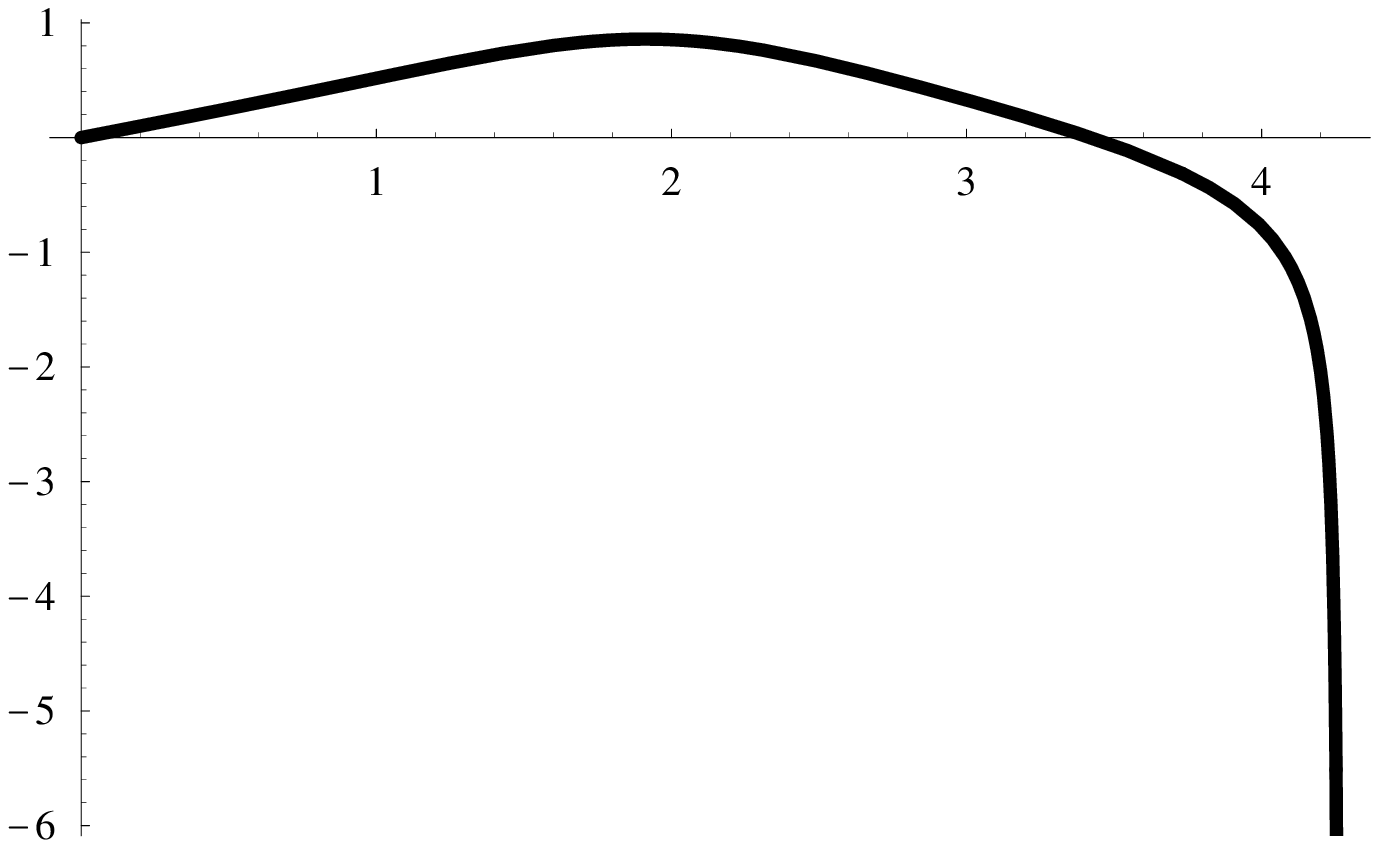)} \\
\multicolumn{2}{l}{\footnotesize (c) $w_0'=0.275$} \\
\parbox{4cm}{\xpeinture 4cm by 2.5cm (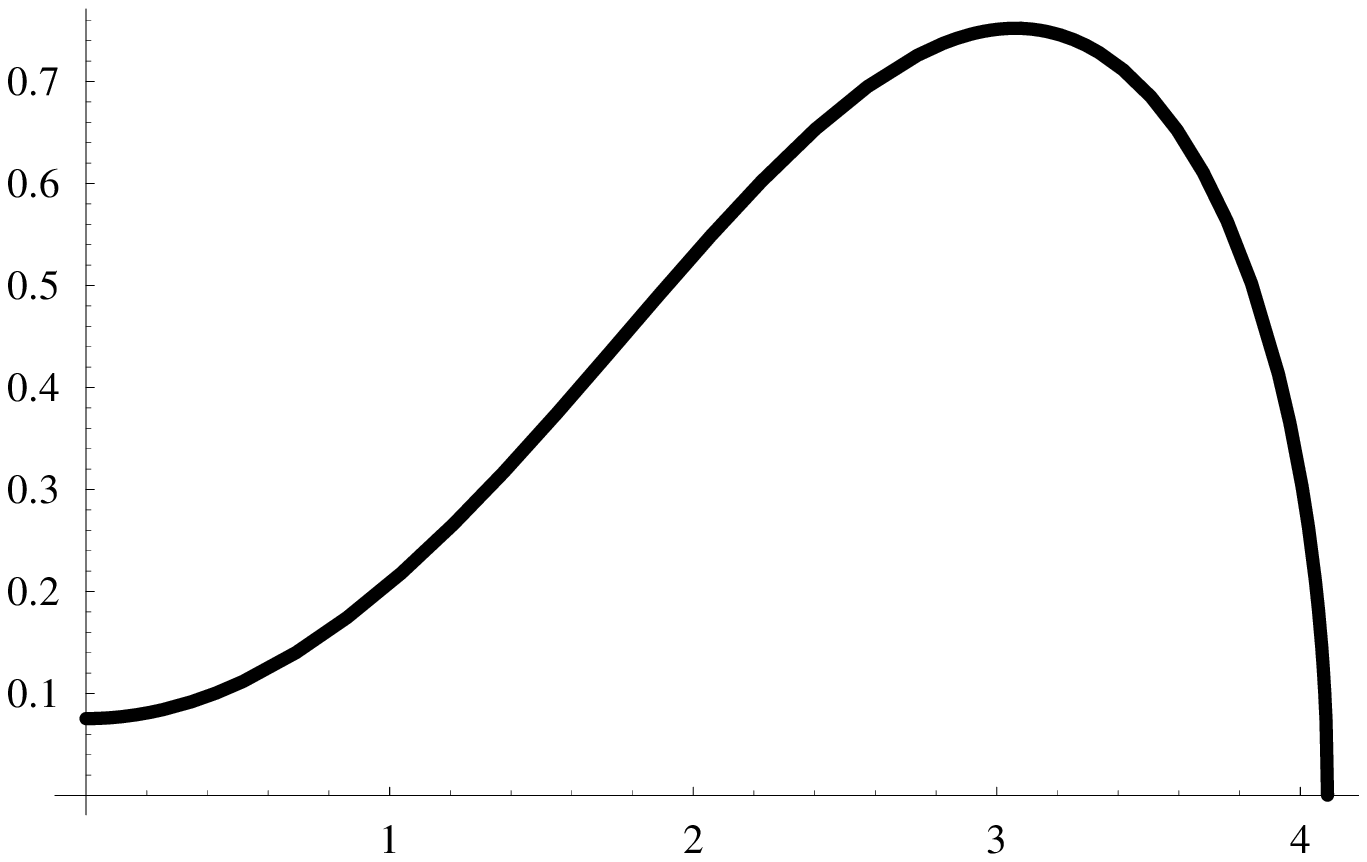)}
&\parbox{4cm}{\xpeinture 4cm by 2.5cm (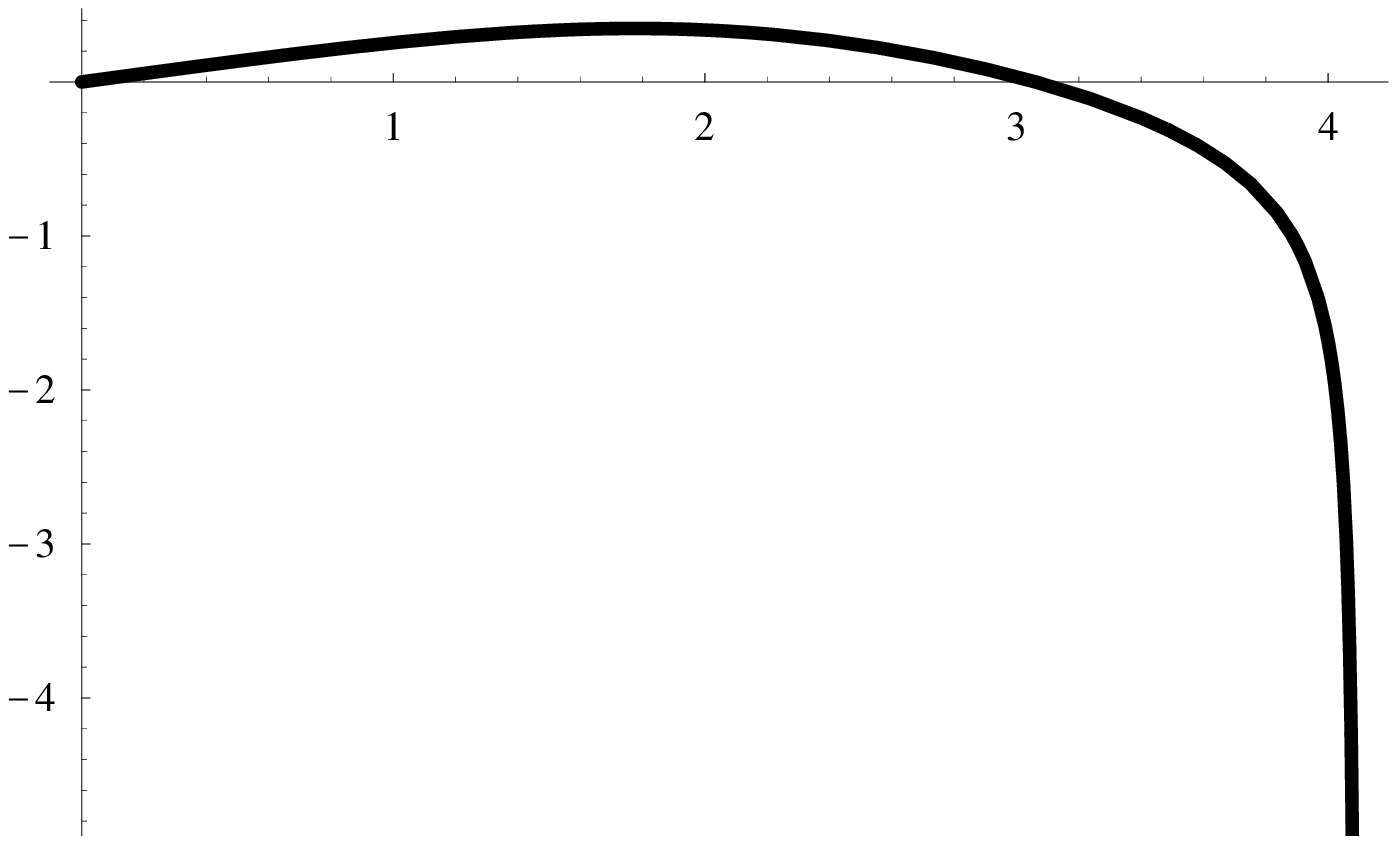)}
 \end{tabular}\\
\hline\hline
\end{tabular}

\section{Equation for Axisymmetric Vesicles}
In this article, we will study axisymmetric solution surface $\Sigma$ of the Helfrich variation problem which has a reflection symmetry by the plane perpendicular to the rotational axis.  If the rotational axis is labelled $z$-axis and the plane of reflection $xy$-plane, then the surface $\Sigma$ can be obtained by revolving a radial curve about the $z$-axis on the upper half plane and reflecting it to the lower half.
\begin{center}
\parbox{8cm}{\xpeinture 7cm by 4.5cm (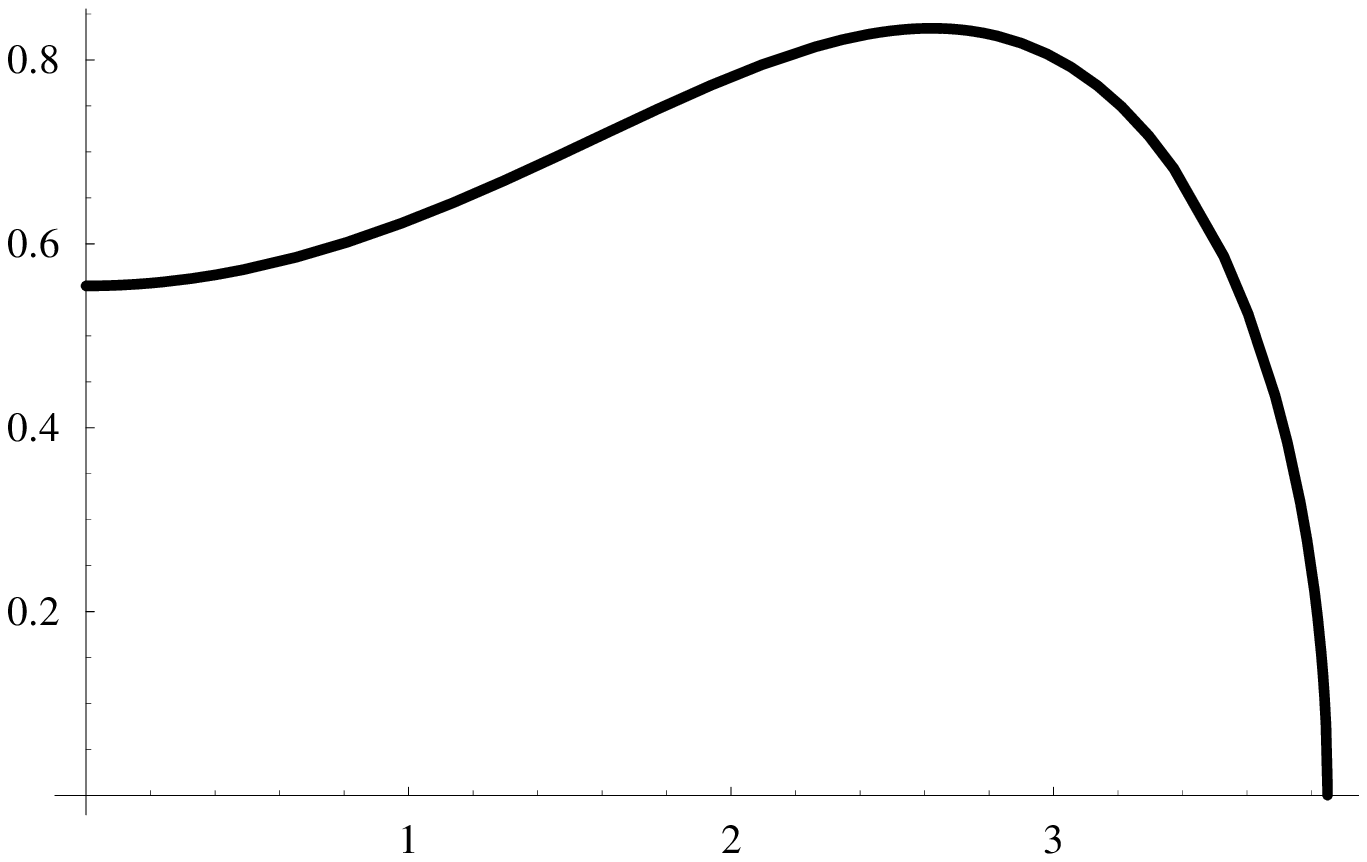)
\caption{A cross-section of a biconcave axisymmetric surface}
\caption{(with $c_0=1$, $\tlambda=0.25$, $\tp=1$)}
}
\end{center}
In other words, one may parametrize the upper part of $\Sigma$ by
$\bX = (r\cos\theta, r\sin\theta, z(r)),$
with a function $z(r)>0$ defined for $r$ in some interval $[0,r_\infty]$, where $r_\infty$ is the radius of the ``equator" that is determined by the surface.
The natural boundary conditions are $z(0)>0$ and $z(r_\infty)=0$.  The rotation and reflection symmetries of the surface then impose the conditions that $z'(0)=0$ and $z'(r)\to -\infty$ as $r\to r_\infty$.  Finally, to model for a biconcave surface, we also need to assume that $z''(0)>0$ in order to have a solution that is concave at the center.  Moreover, the biconcave shape of the surface also confines the graph of $z$ to have a unique point of inflection.  This is equivalent to require that $z'$ has only a unique maximum and no other critical point.  A cross-section of the upper part of $\Sigma$ is shown in the picture.

\subsection*{Equation and Conditions}
Usually, the Helfrich shape equation of axisymmetrc vesicles is written in term of the angle $\psi$ between the surface tangent and the plane perpendicular to rotational axis 
\cite{Helfrich1976,Julicher-Seifert,Zheng-Liu}. However, it is convenient for our discussion to rewrite the equation into a equation on the derivative of the graph $w(r)=z'(r)=\tan \psi$. Then the Helfrich shape equation of axisymmetrc vesicles becomes an equation for $w$, which is
\begin{equation} \label{VEqn1}
\begin{aligned}
\frac{2r}{(1+w^2)^{5/2}}{w''} =  &\frac{5rw}{(1+w^2)^{7/2}}{w'}^2 - \frac{2{w'}}{(1+w^2)^{5/2}} \\
&\qquad {} + \frac{2w+w^3}{r(1+w^2)^{3/2}} + \frac{2c_0 w^2}{1+w^2} + \frac{(c_0^2+\tlambda)r w}{(1+w^2)^{1/2}} - \frac{\tp r^2}{2}.
\end{aligned}
\end{equation}
We are going to study the solution $w(r)$ to this equation with initial choice $w(0)=0$ and $w'(0)={w_0'} > 0$. Our goal is to find solution $w$ that also satisfies the end point condition  that $w(r)\to -\infty$ as $r\to r_\infty$ and an integral condition that $-\infty \ne \displaystyle \int_0^{r_\infty} w\d r < 0$.  The integral condition is equivalent to $z(r_\infty) = 0 < z(0) < \infty$ which ensures that $z(r)$ together with its reflection produce a closed axisymmetric surface without self-intersection.  Furthermore, $w$ is required to have a unique local maximum and no other critical points.  A typical picture for the graph of such $w$ is shown.
\begin{center}
\mbox{\xpeinture 7cm by 4.5cm (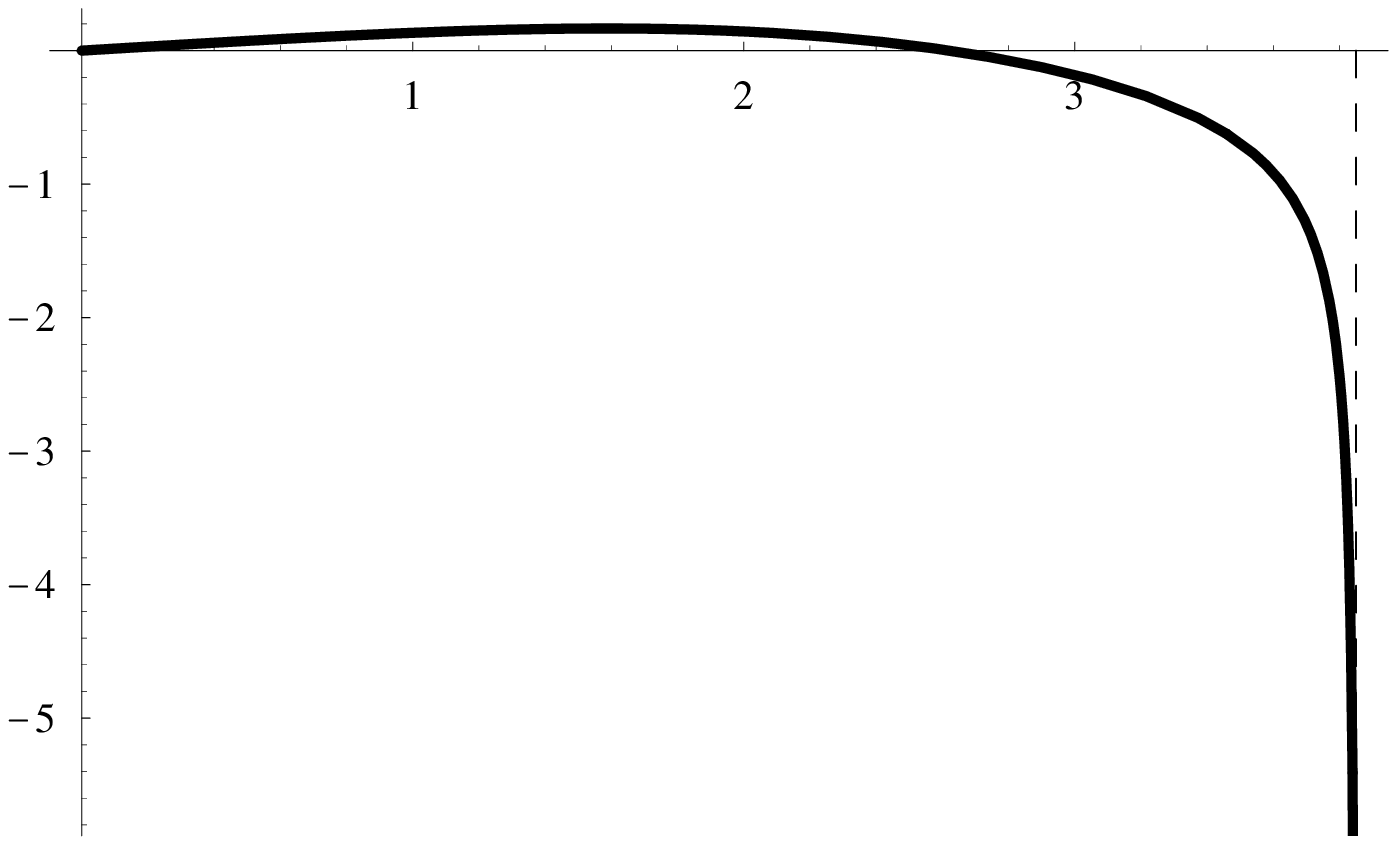)}\\
\parbox{8cm}{{}\hfill{\footnotesize A typically expected graph of $w = z'$}\hfill\hbox{}}
\end{center}

There are two useful ways of grouping the lower order terms in the equation.  A careful study of these lower order terms reveals the qualitative behaviors of the solution $w$.  One of the grouping involves the polynomial $Q(t)$ that we have mentioned in the introduction. 

Let $Q$ be the cubic polynomial
$
Q(t) = t^3 + 2c_0 t^2 + (c_0^2+\tlambda)t - \dfrac{\tp}{2}
$
and $R(t)$ be the quadratic part of $Q(t)$, that is, $R(t) = Q(t) - t^3$.
With this notation, after multiplying with $r {w'}$, the equation~(\ref{VEqn1}) can be written as follow,
\begin{align}
\left[ \frac{r^2 {w'}^2}{(1+w^2)^{5/2}} \right]' &= \left[ \frac{w^2}{(1+w^2)^{1/2}} \right]' + r^3 {w'} R(\kappa(r));\quad\text{or} \label{VEqn2}\tag{\ref{VEqn1}a} \\
\left[ \frac{r^2 {w'}^2}{(1+w^2)^{5/2}} \right]' &= \left[ \frac{-2}{\sqrt{1+w^2}} \right]' + r^3 {w'} Q(\kappa(r)), \label{VEqn3}\tag{\ref{VEqn1}b}
\end{align}
where $\kappa(r) = \dfrac{w}{r\sqrt{1+w^2}}$  is the principal curvature of $\Sigma$ in the meridinal direction.  In other words, the lower order terms capture the changes of the meridinal curvature in the differential equation.
Note that if $c_0 = \tlambda = \tp = 0$, the solution to equation (\ref{VEqn2}) corresponds to the situation that $z=z(r)$ is a circular arc with radius $1/{w_0'}$.  This is exactly the case when the meridinal curvature is a constant.

The initial choice ${w_0'}$ is actually the meridinal curvature at the center of $\Sigma$, i.e.,
$$
\kappa(r) = \frac{w}{r\sqrt{1+w^2}} \to {w_0'}, \qquad \text{as $r\to 0$.}
$$
Moreover, for both $Q$ and $R$, $Q(0) = R(0) = -\tp/2 < 0$.  Therefore, if ${w_0'}$ is less than the smallest positive root of $Q(t)$, then both $Q({w_0'}) < 0$ and $R({w_0'}) < 0$.  Our analysis (in \S 3) shows that these negativity conditions are essential for the meridinal curvature $\kappa(r)$ to change sign.  We will also demonstrate that, for any set of parameters satisfying our condition, a solution $w(r)$ with $w_0'$ small enough always satisfies all of our requirements.  In fact, if exact values of the parameters are given, one can always calculate the suitable value of $w_0'$ numerically.

\section{Necessary conditions for the formation of biconcave solution}
It turns out that the principal curvature of the vesicle $\Sigma$ in the meridinal direction, which we denoted by $\kappa(r) = \dfrac{w}{r\sqrt{1+w^2}}$, is the most basic quantity in our analysis.  The equation (\ref{VEqn3}) implies that if $c_0>0$ and the initial curvature at the center $w_0'$ is large, in the sense that $R(w_0') > 0$, then $\kappa(r)$ increases and a biconcave shape cannot be formed. So a necessary condition for formation of axisymmetric solution of biconcave shape in the case that $c_0>0$ is
$$
R(w_0')=2c_0{w_0'}^2+ (c_0^2 + \tlambda)w_0' - \frac{\tp}{2}<0.
$$

The following pictures show the behaviors of $\kappa(r)$ for two different values of $w_0'$ with $c_0=1$, $\tlambda=0.25$, and $\tp=1$.  For these parameter values, $R(0.277124)=0$.  The one on the left hand side is taken with $w_0'=0.278$ so the necessary condition is not satisfied. In this case, $\kappa$ blows-up at finite distance from the rotational axis so the corresponding solution will not lead to a closed surface of the desired biconcave shape.  The picture on the right hand side is taken with $w_0'=0.276$.  In this case, the necessary condition is satisfied, but $\kappa$ only decays gradually and the solution $w(r)$ still does not give a biconcave surface.  Numerically, we see that $w(r)\to -\infty$ as $r\to r_\infty$ but the total area is positive, i.e., the surface together with its reflection does not close up to form a closed surface in $\R^3$ without self-intersection.  However, we can only show that $R(w_0')<0$ implies the decrease in $\kappa(r)$ without knowing whether it implies the sign change of $\kappa(r)$. So, it is not sure whether $R(w_0')<0$ is sufficient for the formation of closed surface of biconcave shape, a situation in which $\kappa(r)$ has to change from positive to negative.
\begin{center}
\parbox{7cm}{\ypeinture 7cm by 5cm (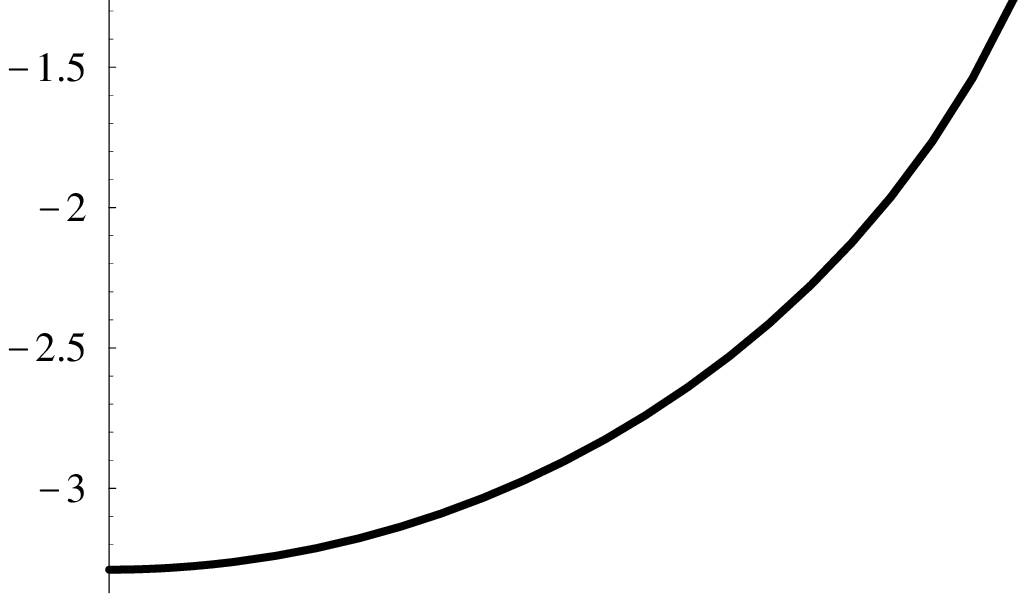) \caption{Graph of $z(r)$ for a convex surface}
\caption{where $R(w_0')>0$}
\caption{Below: the corresponding graph of $w=z'(r)$}
} \hfil
\parbox{7cm}{\ypeinture 7cm by 5.5cm (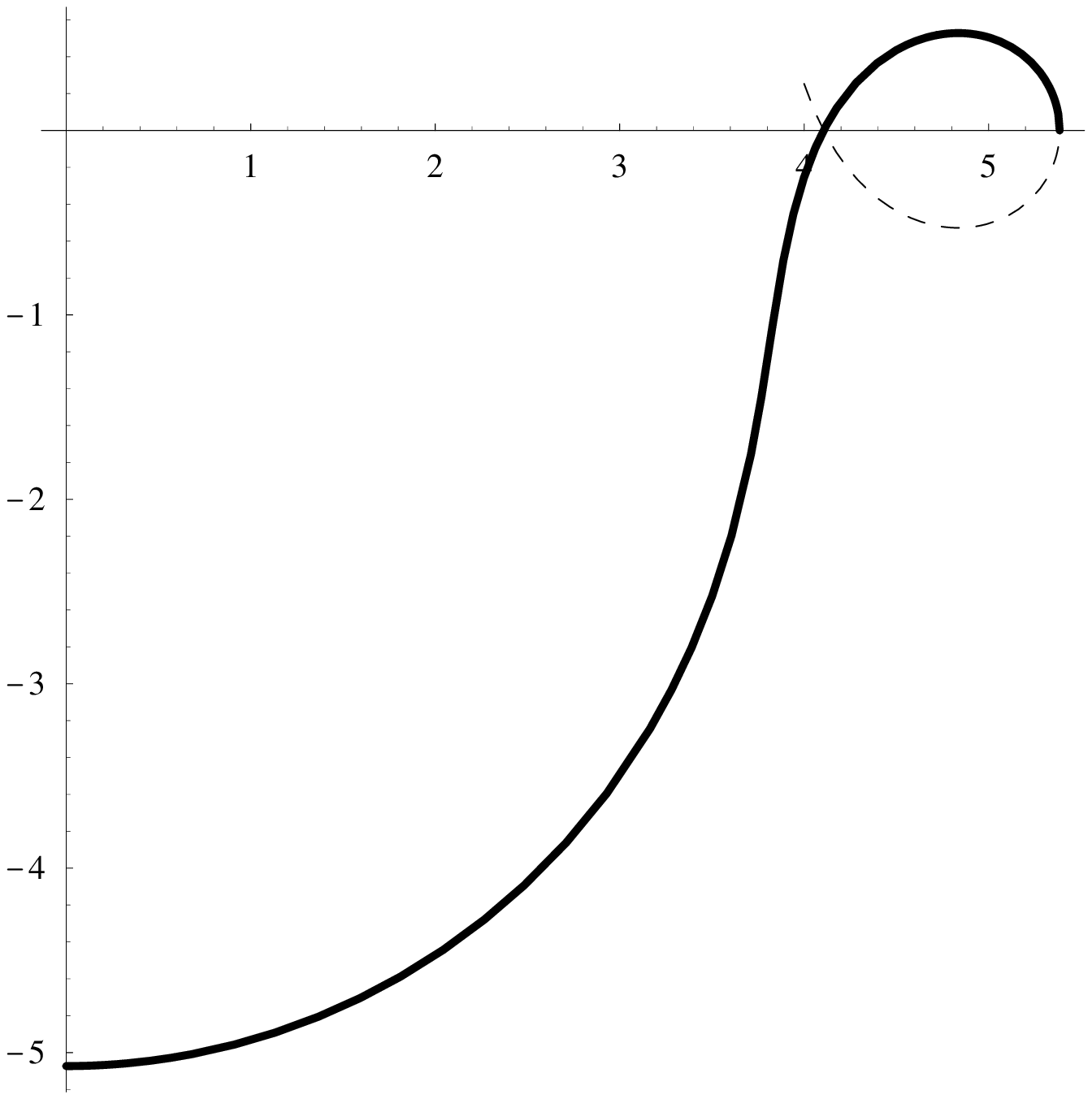) \caption{Graph of $z(r)$ leading to a surface}
\caption{with self-intersection where $R(w_0')<0$}
\caption{Below: the corresponding graph of $w=z'(r)$}
}\\
\mbox{\ypeinture 7cm by 4cm (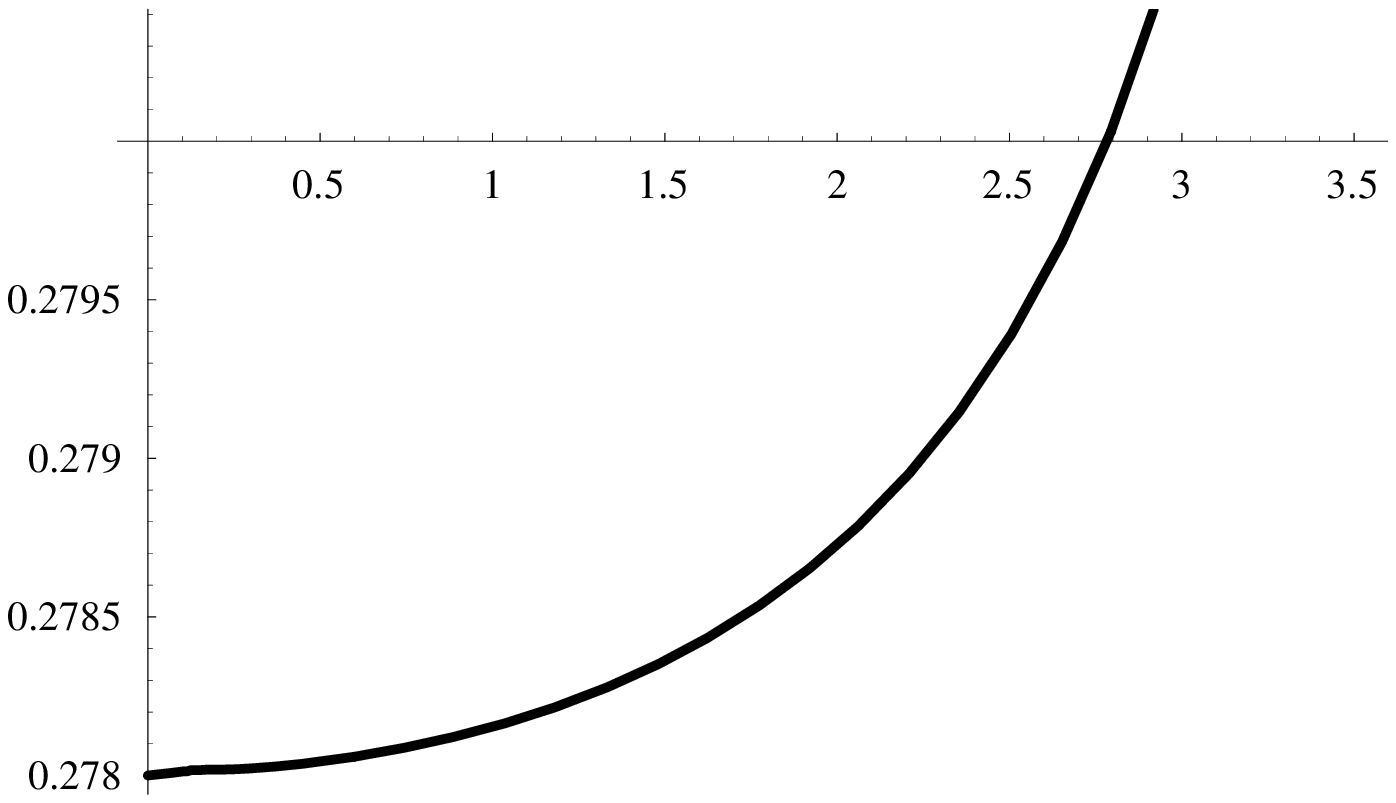)} \hfil
\mbox{\ypeinture 7cm by 4cm (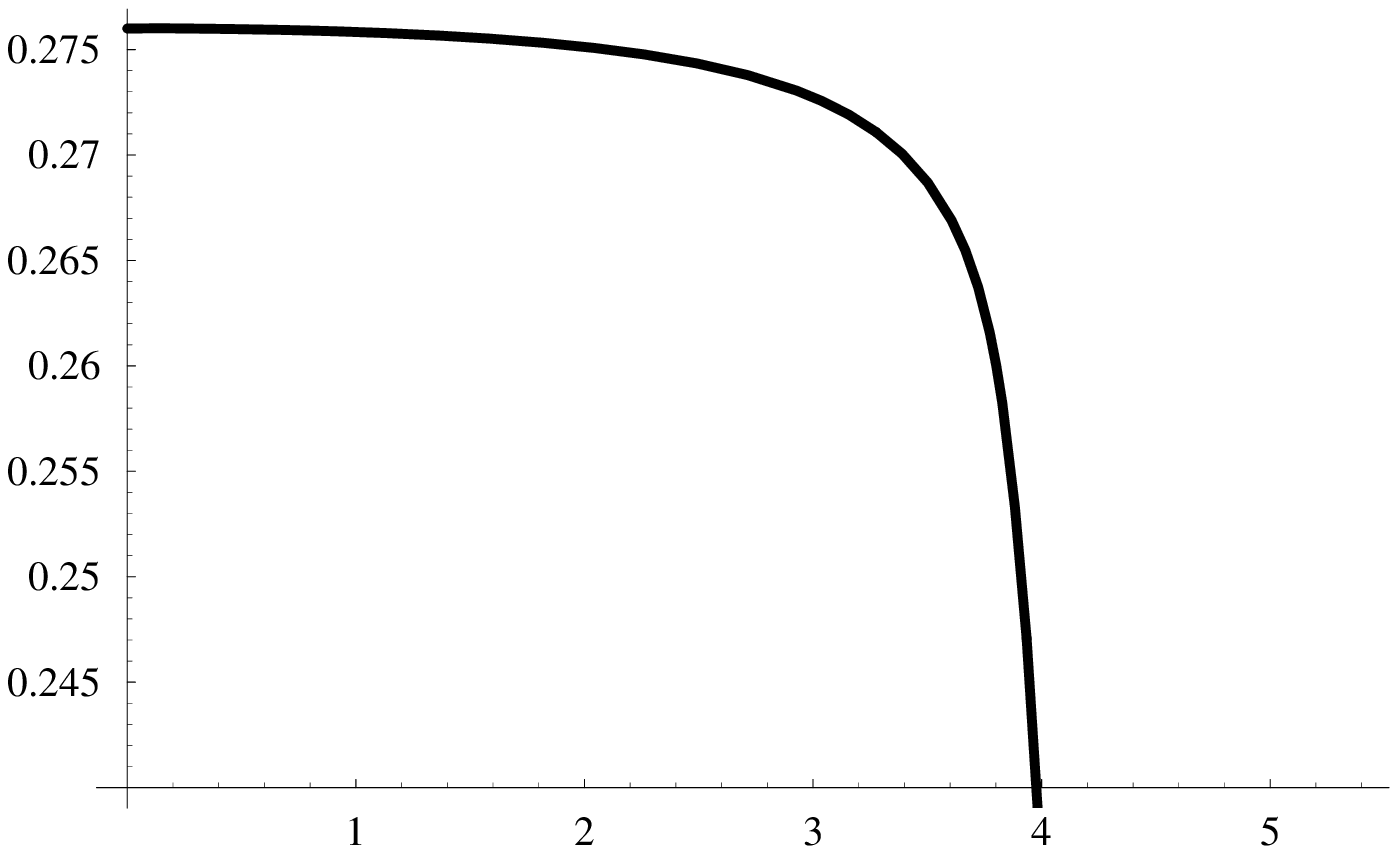)}
\end{center}

In order to obtained an axisymmetric solution with reflection symmetry via a solution of ordinary differential equation, one also needs to verify the first variation is zero with respect to any variation near the plane of reflection. It turns out that this is always the case for the solution we are going to study in the next section.  Moreover, the analysis shows that the radius $r_\infty$ of the ``equator" and the Guassian curvature $K(r_\infty)$ of the vesicle at any point on the ``equator" must satisfy the following relation,
$$
K(r_\infty)^2 = \frac{-1}{r_\infty} Q\left(\frac{-1}{r_\infty}\right).
$$
This relation is easily verified for the case that $c_0=\tlambda=\tp=0$. In this case, we have $Q(t)=t^3$ and $K(r_\infty)^2=\dfrac{1}{r_\infty^4}$. This is compatible with the fact that the solution surface is a round sphere.

\section{Sufficient condition for the formation of biconcave solution}
In the preceding section, although that $R(w_0')<0$ is known to be a necessary condition for the case that $c_0>0$, it may not be sufficient.  Hence, we need to impose a stronger condition that all roots of $Q(t)$ are positive and require $w_0'$ to be small enough. Unfortunately, we do not have a general formula for the threshold value of $w_0'$. 

The smallness assumption on $w_0'$ implies that not only $R(w_0')<0$ but also $Q(w_0')<0$. In this case, the meridinal curvature changes sign.  And in terms of $w(r)$, though it is initially positive, it becomes negative when $r$ reaches a certain value $r_0$. Note that for different initial value $w_0'$, the value of $r_0$ is different. Although the precise relationship is unknown, we know that $r_0$ is comparable to ${w_0'}^{1/2}$ when $w_0'$ is small. More precisely, we have the following limiting inequalities
$$
\frac{16}{3\tp}\le \lim_{w_0'\to 0}\frac{r_0^2}{w_0'} \leq \frac{16}{\tp}.
$$
In fact, we discover that, for any point of inflection $r_c$ of the graph of $z(r)$ with $\kappa(r_c)>0$,
$$
\frac{16}{3\tp}\le \lim_{w_0'\to 0}\frac{r_c^2}{w_0'}\le\lim_{w_0'\to 0}\frac{r_0^2}{w_0'}\leq \frac{16}{\tp}.
$$
As a consequence of these inequalities, the solution will have a unique point of inflection of the graph of $z(r)$.
Moreover, for the case that all roots of $Q(t)$ are positive (see \S 2), the function $w(r)$ blows down monotonically to $-\infty$ in a finite distance from the rotational axis after it becomes negative.  

The following are pictures of $z(r)$ and $w(r)$ for $c_0=1$, $\tlambda=0.25$, $\tp=1$, and $w_0'=0.26$. This set of parameters satisfies our condition. In fact, $Q(t)$ has a unique positive root at 0.268828. So, it is clear that $Q(w_0')<0$ and the function $w(r)$ goes to $-\infty$ at $r_\infty \approx 5.39215$. However, $w_0'=0.26$ is not small enough to give a solution $w(r)$ with total negative area and hence it does not give a vesicle of the required biconcave shape.

\begin{center}
\mbox{\xpeinture 7cm by 4.5cm (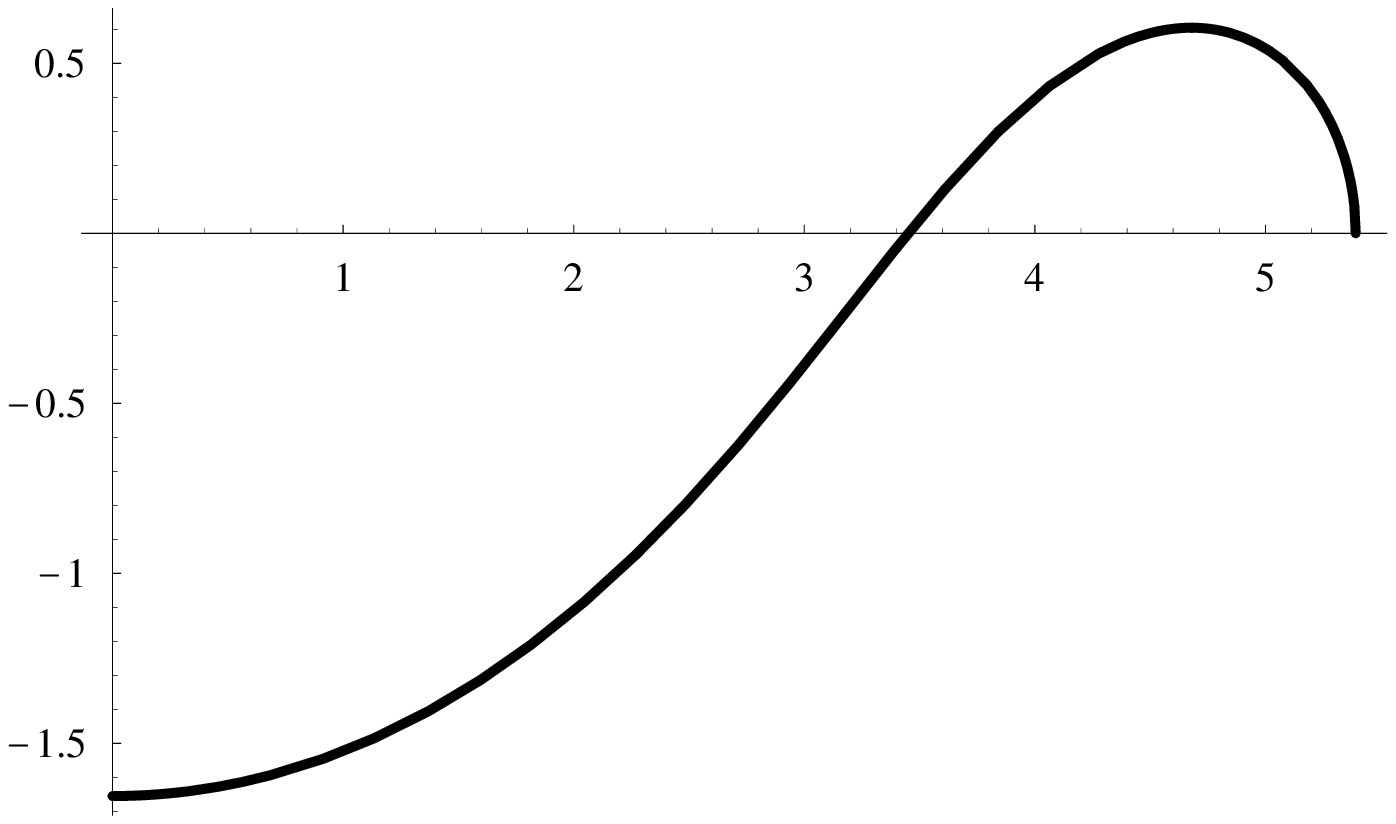)} \hfil
\mbox{\xpeinture 7cm by 4.5cm (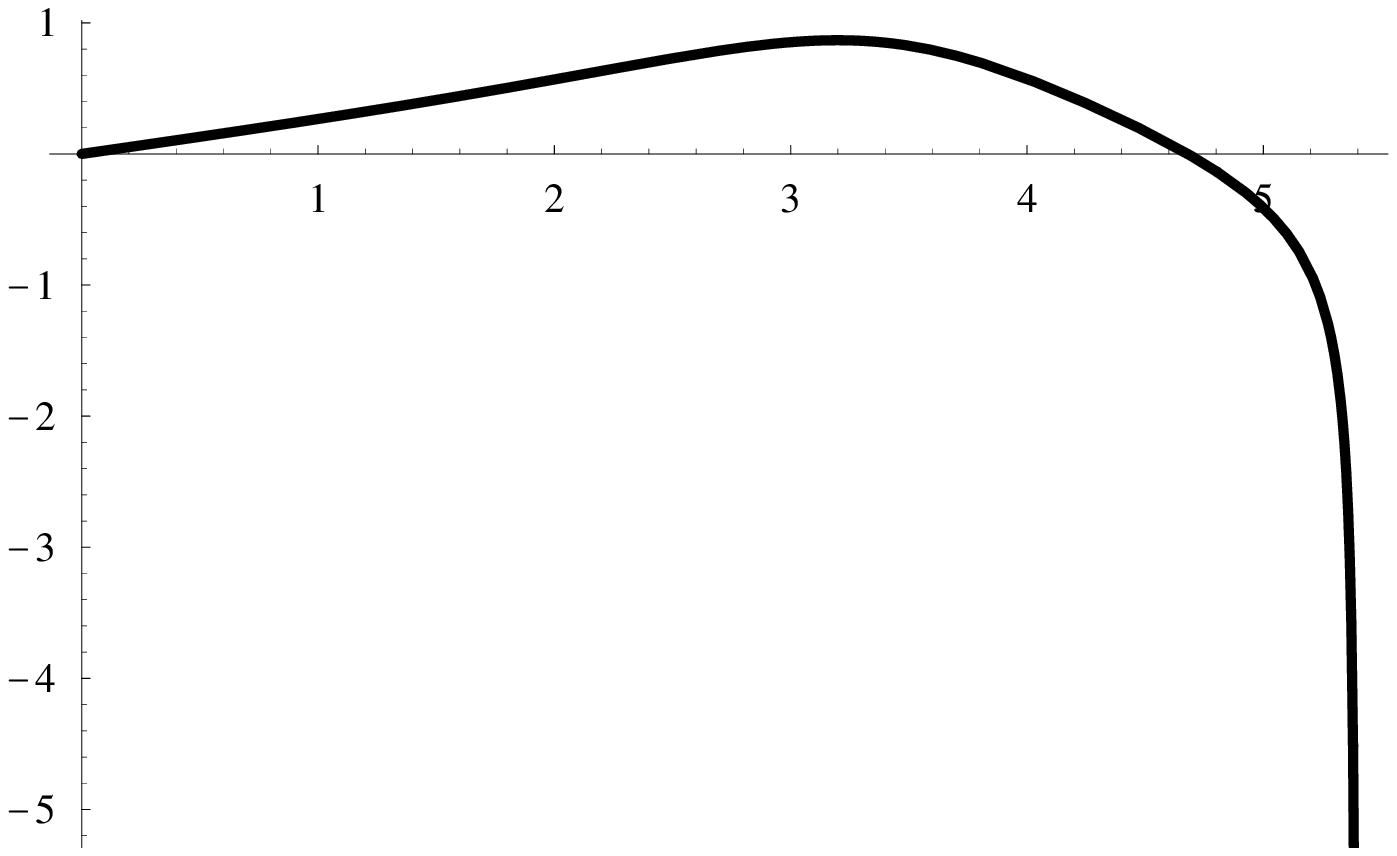)} \\
\mbox{\footnotesize A solution $z$ which does not give an embedded surface although $w=z'$ blows down to $-\infty$}
\end{center}

However, if we choose $w_0' = 0.2063860$ and $w_0' = 0.15$ for the same set of parameters $c_0=1$, $\tlambda=0.25$, and $\tp=1$, then in each case, the total area under the graph of $w(r)$ is nonpositive. Hence, each of the graph of $z(r)$ can be reflected and rotated to form a vesicle of biconcave shape. The following pictures are the function $z(r)$ and $w(r)$ for these values of $w_0'$ respectively.
\begin{center}
\mbox{\xpeinture 7cm by 4.5cm (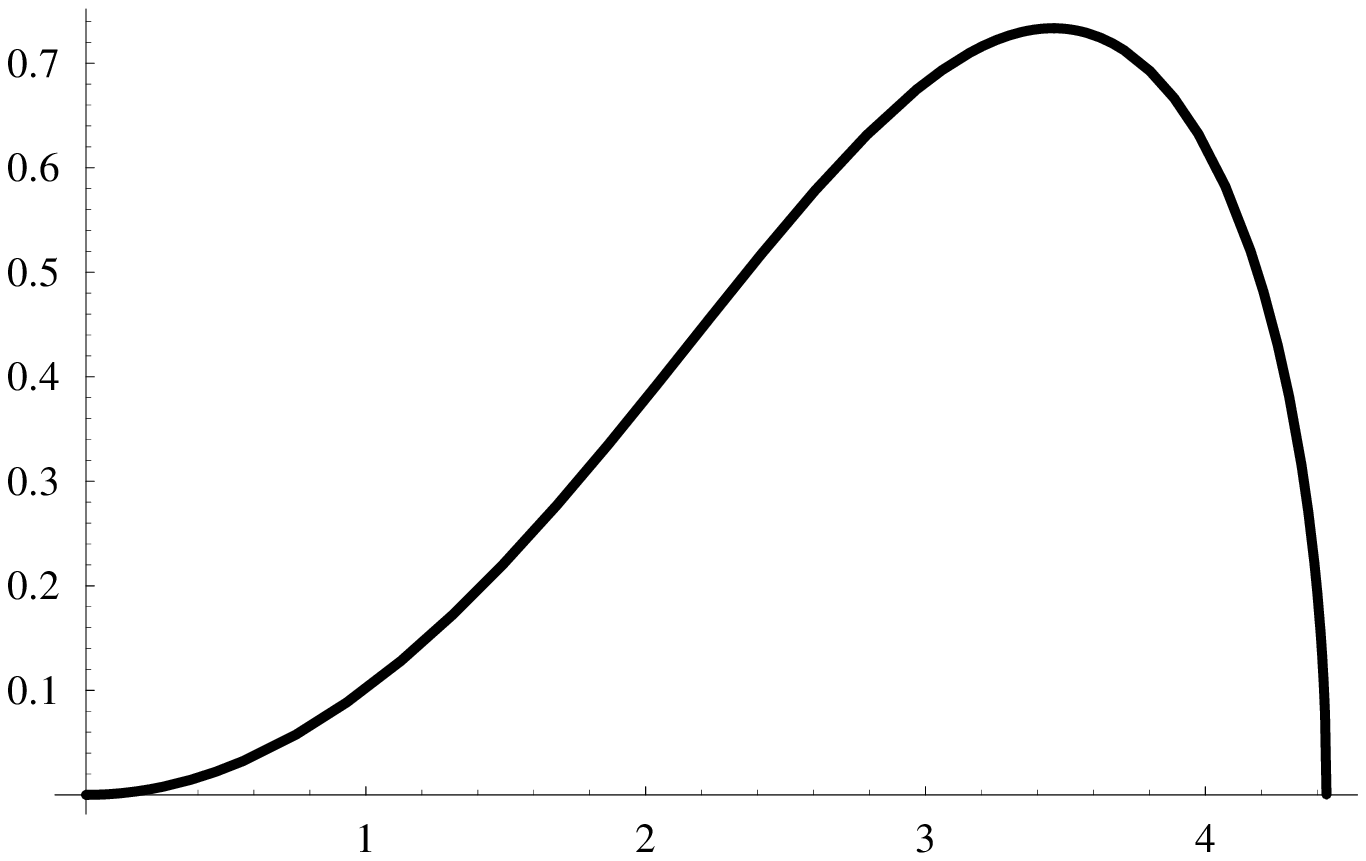)} \hfil
\mbox{\xpeinture 7cm by 4.5cm (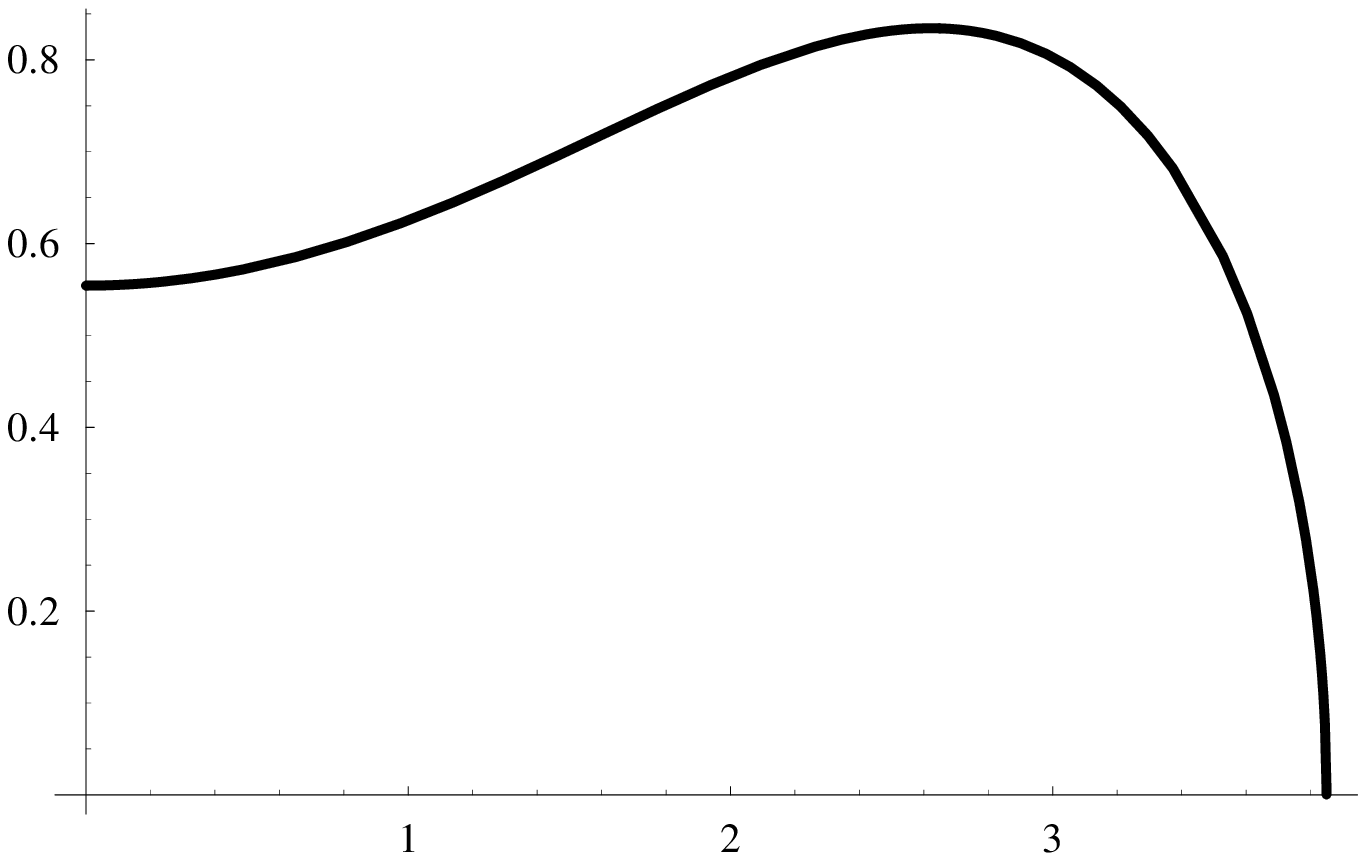)} \\
\mbox{\footnotesize Graphs of $z(r)$ giving rise to biconcave surfaces and their $z'$ below} \\
\mbox{\xpeinture 7cm by 4.5cm (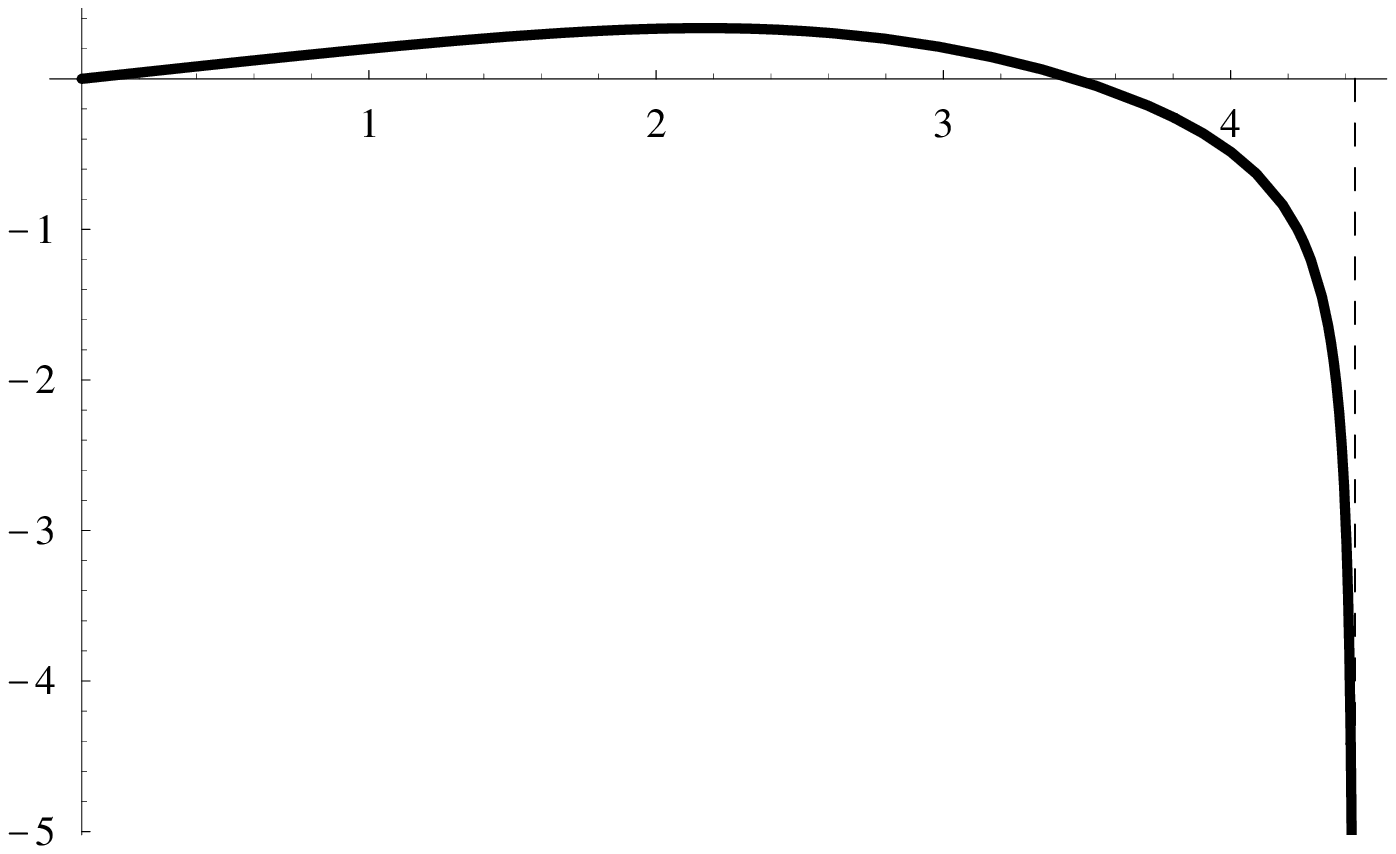)} \hfil
\mbox{\xpeinture 7cm by 4.5cm (wareaneg.ps)}
\end{center}
The mathematical proof of this result involves more delicate study of the equation~(\ref{VEqn3}) together with sharp estimates of $r_0$, $r_\infty$, the rate of change of $w(r)$ at $r_0$, and the integral of $w(r)$.  Please refer to the mathematical paper by the authors for the details of the analysis.


\section{Discussion}
Finally, we would like to remark on other solutions which do not correspond to biconcave surfaces.  Firstly, when $c_0^2+\tlambda<0$, one may still obtain solution for nonconvex surface even $w_0'$ is not very small.
\begin{center}
\mbox{\xpeinture 7cm by 4.3cm (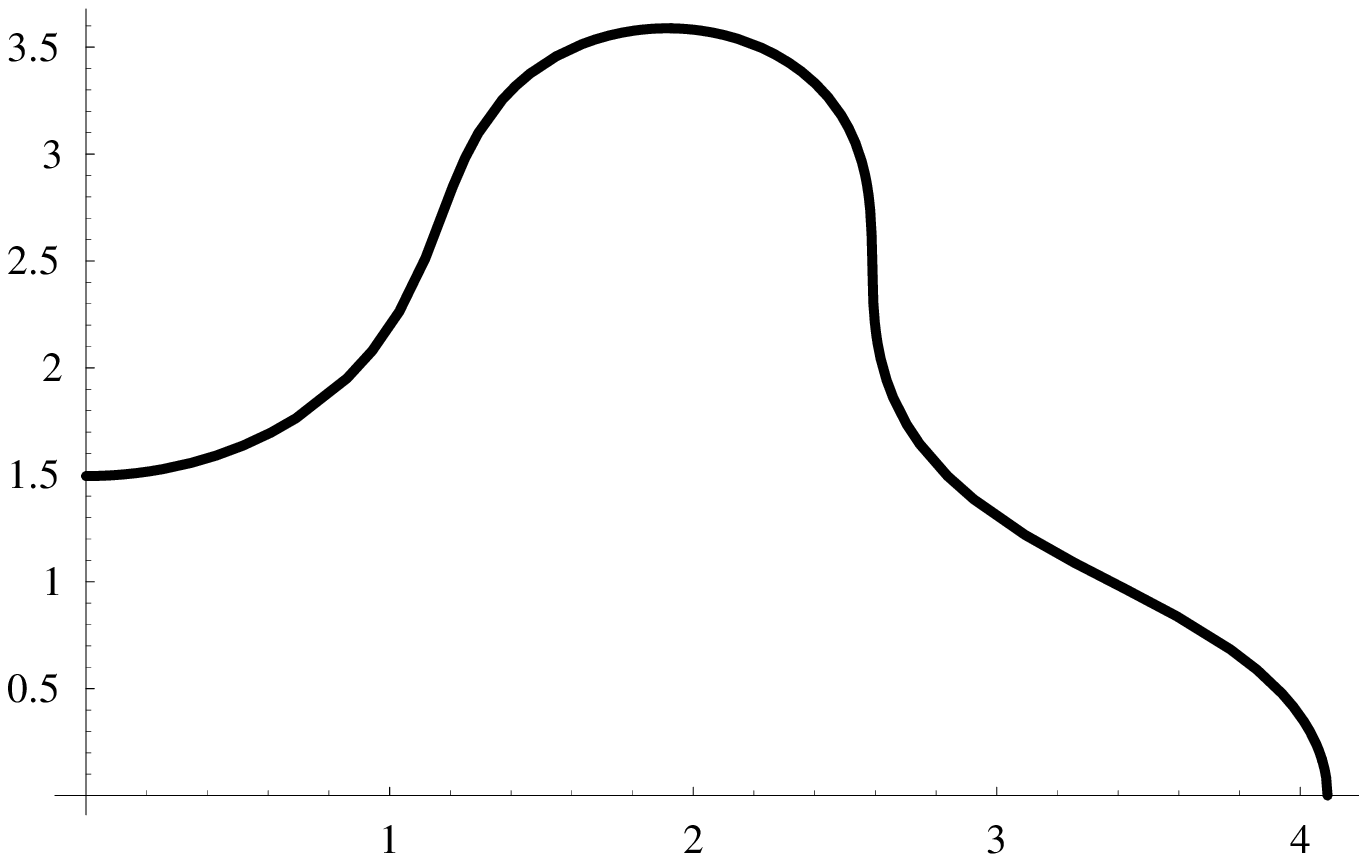)}\\
\mbox{\footnotesize A solution which gives a nonconvex surface}
\end{center}
The above picture is a solution for $c_0=1$, $\tlambda=-3$, $\tp=2$ with $w_0'=2$.  Note that $w_0'$ is greater than the positive roots of $R(t)$ and $Q(t)$.  Moreover, $Q(t)$ has negative roots and thus this is beyond the situation that we discuss in previous sections.  In this case, after $w=z'$ becomes negative, it may not decrease all the way to negative infinity, which is why a second ``petal'' may form.

The other type is a rotational solution which is not symmetric under reflection in $xy$-plane \cite{Helfrich1976,Helfrich1977}.  The following is a solution with $c_0 = -1$, $\tlambda = -1$ and $\tp = 1$.
\begin{center}
\mbox{\xpeinture 5.5cm by 5.5cm (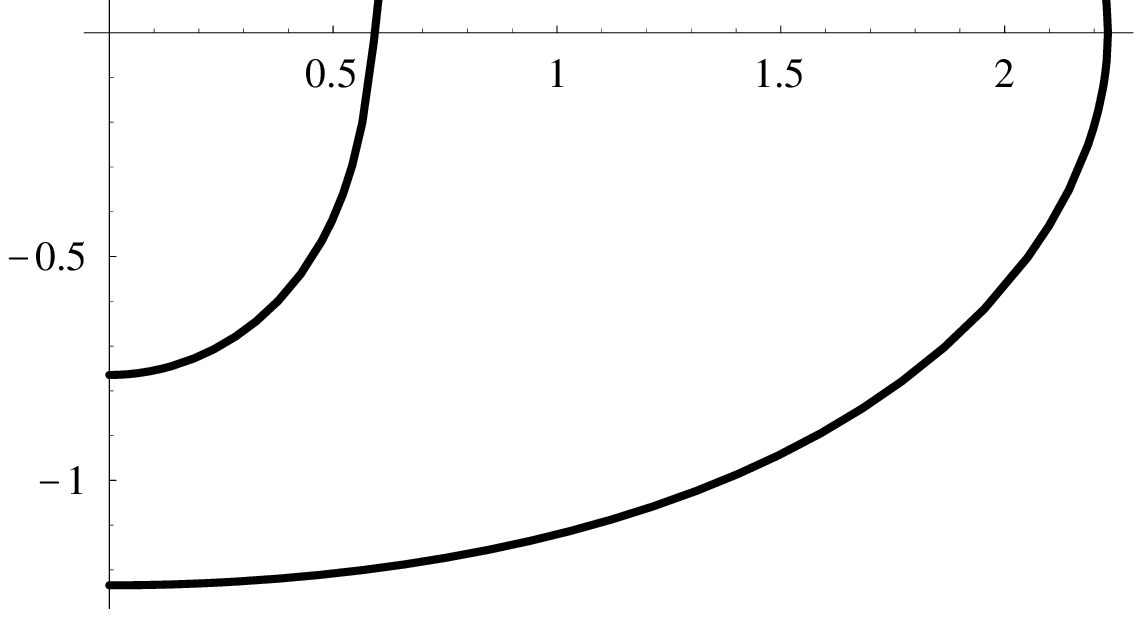)}\\
\mbox{\footnotesize A solution which has no reflection symmetry}
\end{center}
It can be observed that the upper part of this solution satisfies our equation~(\ref{VEqn1}) but not the area condition.  Although the reflected graph is also a solution locally, it does not give rise to a surface without self-intersection.  For this combination of parameters, there is a bifurcation of solution at $r_\infty$ to close up by an asymmetric surface.  From our numerical study, it should be remarked that not all solutions in the upper quadrant may be closed up by an asymmetric counterpart.  For example, if the upper part satisfies the area condition, the only possible lower part is the reflection symmetric one.  This suggests a possible uniqueness of biconcave solution.  Furthermore, we surmise that for $c_0 > 0$, $\tlambda > 0$ and $\tp > 0$, solution of the asymmetric type never occurs.  The local minimum of $Q(t)$ for $t < 0$ may be the underlying obstruction of such happening.  However, at the moment, there is not yet a mathematical proof.


\begin{thebibliography}{WWW}
\bibitem[DH]{Helfrich1976}
H.~J.~Deuling and W.~Helfrich.
\newblock{\em
The Curvature Elasticity of Fluid Membranes: A Catalogue of Vesicle Shapes.\/
}
\newblock{
Journal de Physique. Tome 37 (Nov 1976), 1335--1345.}

\bibitem[HDH]{Helfrich1977}
W.~Harbich, H.~J.~Deuling, and W.~Helfrich.
\newblock{\em
Optical Observation of Rotationally Symmetric Lecithin Vesicle Shapes.\/
}
\newblock{
Le Journal de Physique.  Tome 38 (Jun 1977), 727--729.
}

\bibitem[H1]{Helfrich1973}
W.~Helfrich.
\newblock{\em
Elastic Properties of Lipid Bilayers: Theory and Possible Experiments.\/}
\newblock{
Z.~Naturforsch.  Teil~C, 28 (1973), 693--703.}

\bibitem[HO]{Hu-OuYang}
J.~Hu and Z.~Ou-Yang.
\newblock{\em
Shape Equations of the Axisymmetric Vesicles.\/
}
\newblock{
Physical Review E.  Vol~47, No~1 (Jan 1993), 461--467.
}

\bibitem[JS]{Julicher-Seifert}
F.~J\"ulicher and U.~Seifert.
\newblock{\em
Shape Equations for Axisymmetric Vesicles: A Clarification.\/
}
\newblock{
Physical Review E.  Vol~49, No~5 (May 1994), 4728--4731.
}

\bibitem[L]{Luke}
Jon Luke.
\newblock{\em
A Method for the Calculation of Vesicle Shapes.\/
}
\newblock{
SIAM Journal of Applied Mathematics.  Vol~42, No~2 (Apr 1982), 333--345.
}

\bibitem[MB]{Mutz-Bensimon}
M.~Mutz and D.~Bensimon.
\newblock{\em
Observation of toroidal vesicles.\/
}
\newblock{
Physical Review A.  Vol~43, No~8 (Apr 1991), 4525--4527.
}

\bibitem[NOO1]{Naito-Okuda-OuYang1}
H.~Naito, M.~Okuda, and Z.~Ou-Yang.
\newblock{\em
Counterexample to some Shape Equations for Axisymmetric Vesicles.\/
}
\newblock{
Physical Review E.  Vol~48, No~3 (Sept 1993), 2303--2307.
}

\bibitem[NOO2]{Naito-Okuda-OuYang2}
H.~Naito, M.~Okuda, and Z.~Ou-Yang.
\newblock{\em
New Solutions to the Helfrich Variation Problem for the Shapes of Lipid Bilayer Vesicles: Beyond Delaunay's Surfaces.\/
}
\newblock{
Physical Review Letters.  Vol~74, No~21 (May 1995), 4345--4348.
}

\bibitem[OH1]{OuYang-Helfrich1987}
Z.~Ou-Yang and W.~Helfrich.
\newblock{\em
Instability and Deformation of a Spherical Vesicle by Pressure.\/
}
\newblock{
Physical Review Letters.  Vol~59, No~21 (Nov 1987), 2486--2488.
}

\bibitem[OH1]{OuYang-Helfrich1989}
Z.~Ou-Yang and W.~Helfrich.
\newblock{\em
Bending energy of vesicle membranes: General expressions for the first, second, and third variation of the shape energy and applications to spheres and cylinders.\/
}
\newblock{
Physical Review A.  Vol~39, No~10 (May 1989), 5280--5288.
}

\bibitem[OY]{OuYang}
Z.~Ou-Yang.
\newblock{\em
Anchor ring-vesicle membranes.\/}
\newblock{
Physical Review A.  Vol~41, No~8 (April 1990), 4517--4520.}

\bibitem[S]{Seifert}
U.~Seifert.
\newblock{\em
Vesicles of Toroidal Topology.\/
}
\newblock{
Physical Review Letters.  Vol~66, No~18 (May 1991), 2404--2407.
}

\bibitem[W]{Willmore}
T.~J.~Willmore.
\newblock{\em
Riemannian Geometry.\/}
\newblock{
Clarendon Press, Oxford, 1993.}

\bibitem[ZL]{Zheng-Liu}
W.~Zheng and J.~Liu.
\newblock{\em
Helfrich Shape Equation for Axisymmetric Vesicles as a First Integral.\/
}
\newblock{
Physical Review E.  Vol~48, No~4 (Oct 1993), 2856--2860.
}


\end{thebibliography}
\end{document}